\documentclass[useAMS, usenatbib, usegraphicx]{mn2e}

\usepackage{times}

\newcommand{\ion}[2]{\mbox{{#1}\,{\sc {#2}}}}
\newcommand{\kms}{\,\mbox{km s$^{-1}$}}
\def\apj{ApJ}
\def\aap{A\&A}
\def\mnras{MNRAS}
\def\apjl{ApJ}
\def\araa{ARA\&A}
\def\nat{Nature}
\def\aaps{A\&AS}
\def\apjs{ApJS}
\def\pasp{PASP}
\def\aj{AJ}
\title[The Molecular Hydrogen Accretion Disc in NGC~1275]{Kinematics and Excitation of the Molecular Hydrogen Accretion Disc in NGC~1275}
\author[J. Scharw\"achter et al.]{J. Scharw\"achter$^{1}$\thanks{E-mail:
julia@mso.anu.edu.au}, P. J. McGregor$^{1}$, M. A. Dopita$^{1,2,3}$, and T. L. Beck$^{4}$\\
$^{1}$Research School of Astronomy and Astrophysics, The Australian National University, Mount Stromlo Observatory, \\Cotter Road, Weston Creek 2611, Australia\\
$^{2}$Astronomy Department, King Abdulaziz University,
P.O. Box 80203, Jeddah, Saudi Arabia\\
$^{3}$Institute for Astronomy, University of Hawaii, 2680 Woodlawn Drive, Honolulu, HI 96822, USA\\
$^{4}$Space Telescope Science Institute, 3700 San Martin Drive, Baltimore, MD 21218, USA}
\begin{document}

\date{}

\pagerange{\pageref{firstpage}--\pageref{lastpage}} \pubyear{}

\maketitle

\label{firstpage}

\begin{abstract}
We report the results of high spatial and spectral resolution integral-field spectroscopy of the central $\sim 3 \times 3$~arcsec$^2$ of the active galaxy NGC~1275 (Perseus~A), based on observations with the Near-infrared Integral Field Spectrograph (NIFS) and the ALTAIR adaptive-optics system on the Gemini North telescope. 
The circum-nuclear disc in the inner $R\sim 50$~pc of NGC~1275 is seen in both the H$_2$ and [\ion{Fe}{ii}] lines. The disc is interpreted as the outer part of a collisionally-excited turbulent accretion disc. The kinematic major axis of the disc at a position angle of $68^\circ$ is oriented perpendicular to the radio jet. A streamer-like feature to the south-west of the disc, detected in H$_2$ but not in [\ion{Fe}{ii}], is discussed as one of possibly several molecular streamers, presumably falling into the nuclear region. Indications of an ionization structure within the disc are deduced from the \ion{He}{i} and Br$\gamma$ emission lines, which may partially originate from the inner portions of the accretion disc. The kinematics of these two lines agrees with the signature of the circum-nuclear disc, but both lines display a larger central velocity dispersion than the H$_2$ line.
The rovibrational H$_2$ transitions from the core of NGC~1275 are indicative of thermal excitation caused by shocks and agree with excitation temperatures of $\sim 1360$ and $\sim 4290$~K for the lower- and higher-energy H$_2$ transitions, respectively. The data suggest X-ray heating as the dominant excitation mechanism of [\ion{Fe}{ii}] emission in the core, while fast shocks are a possible alternative. The [\ion{Fe}{ii}] lines indicate an electron density of $\sim 4000\ \mathrm{cm^{-3}}$. The H$_2$ disc is modelled using simulated NIFS data cubes of H$_2$ emission from inclined discs in Keplerian rotation around a central mass. Assuming a disc inclination of $45^\circ \pm 10^\circ$, the best-fitting models imply a central mass of $(8 ^{\ + 7}_{\ - 2}) \times 10^8\ \mathrm{M_\odot}$. Taken as a black-hole mass estimate, this value is larger than previous estimates for the black-hole mass in NGC~1275, but is in agreement with the $M$-$\sigma$ relation within the rms scatter. However, the molecular gas mass in the core region is tentatively estimated to be non-negligible, which suggests that the central mass may rather represent an upper limit for the black-hole mass. In comparison to other H$_2$-luminous radio galaxies, we discuss the relative role of jet feedback and accretion in driving shocks and turbulence in the molecular gas component.
\end{abstract}


\section{INTRODUCTION}
\label{intro}

NGC~1275 (also commonly known as Perseus~A, 3C~84, 4C+41.07, or IRAS 03164+4119) is the central galaxy in the Perseus cluster, Abell~426. The cluster Abell~426 shows two distinct velocity components: NGC~1275 ($z=0.017559$) belongs to the low-velocity component, which is associated with velocities of 5200~\kms. The high-velocity component, showing velocities of 8200~\kms\ is located in front of NGC~1275 to the north and north-west of the nucleus \citep{Rubin77}. 
The two velocity components have been interpreted as indicating an ongoing galaxy cluster merger \citep{Unger90}. However, X-ray absorption studies by \citet{Gillmon04} and \citet{Sanders07} indicate that the separation between both velocity components is large. \citet{Gillmon04} suggest that the high-velocity system is not yet merging with NGC~1275 but is still in the process of falling into the cluster. 

The number of classifications for NGC~1275 in the literature --  Fanaroff-Riley~1 radio galaxy (F-R I), BL Lac object, post-merger galaxy, cooling-flow galaxy, peculiar Seyfert~1, and LINER -- reflect the fact that NGC~1275 exhibits an extraordinarily rich set of interstellar physical phenomena. It is a star-forming early-type cD galaxy, harbouring an active nucleus which is a point source of hard X-rays \citep{Kawara93, Prieto96} and a very strong radio source (hence the name Perseus~A). The radio emission on large-scales and parsec-scales shows a pair of jets \citep{Pedlar90, Walker94}. The jets are oriented at a position angle of 160$^{\circ}$ \citep{Pedlar90}. 
Using VLBI measurements of the parsec-scale jet, \citet{Walker94} inferred a likely range of angles to the line of sight of $\theta =30^\circ -55^\circ$. The southern jet is the one approaching us.

NGC~1275 is associated with strong thermal X-ray emission, which extends over several arc minutes around the active nucleus and shows a complex bubble-like morphology \citep{Fabian00,Fabian03a}, as well as with an extended structure of H$\alpha$ filaments \citep{Minkowski57,Lynds70}. These filaments of ionized gas have been studied in detail by \citet{Conselice01}. They are likely to be stabilized by magnetic fields \citep{Fabian08} and display LINER-like spectra which have been analysed by \citet{Sabra00}. \citet{Sabra00} examined a variety of excitation mechanisms -- including shocks, ionization by an AGN, and stellar photoionization -- without finding a fully adequate explanation for the observed spectra. Collisional heating by ionizing particles has recently been investigated in greater detail as a possible explanation for the emission characteristics of the filaments \citep{Ferland09, Fabian11}.

NGC~1275 was originally believed to represent the prototypical cooling-flow galaxy, in which the H$\alpha$ filaments have formed by cooling of the intracluster medium \citep{Fabian94}. But the correlation between the X-ray and H$\alpha$ morphology suggests that the filaments rather consist of cold gas which is uplifted behind the radio bubbles driven by the radio jet \citep{Fabian03b}. \citet{Salome06} and \citet{Salome11} have shown that there is also a tight correlation between the extended CO line emission, the H$\alpha$ filaments, the extended X-ray emission, and the large radio lobes mapped by \citet{Pedlar90}. Likewise, \citet{Lim12} found a close correspondence between the H$\alpha$ filaments and the extended H$_2$ \mbox{1-0} S(1) emission. A fraction of the molecular gas and much of the thermal X-ray emission are confined in shells around bubbles of non-thermal synchrotron-emitting plasma. This is clear evidence of the importance of the radio lobes in compressing and cooling the ambient galactic medium. This ambient galactic medium can be derived from either shock-cooled thermal X-ray plasma or from the debris of a recent galactic merger \citep{Conselice01}. 

The total mass of molecular gas in NGC~1275 is large. Based on \mbox{CO(2-1)} observations 
of the nuclear and extended gas, \citet{Salome06} measured a total mass of cold H$_2$ of $4\times 10^{10}\ \mathrm{M_\odot}$, using the standard CO to molecular hydrogen conversion factor. About $3\times 10^{9}\ \mathrm{M_\odot}$ of this molecular gas are contained in the inner $R\sim 3$~kpc, a mass that is comparable with the total mass of H$_2$ in the Milky Way. 
Most of the emission arises in an extended east-west structure passing through the nucleus.
\citet{Salome06} found that the CO emission in the central 8~kpc diameter region does not show any evidence of large-scale rotation. However, the CO line profile in the central 4~kpc diameter region is double-peaked, which may indicate inflows, outflows, or a rotating disc \citep{Salome11}.

The source of the molecular hydrogen in the nuclear region of NGC~1275 is thought to be turbulent infalling gas. Whether this gas is carried into the central regions by a galactic merger or a cooling flow is unclear \citep[e.g.][]{Salome06}. 
Interferometric observations of the CO emission in the central 10~kpc by \citet{Lim08} revealed that most of the gas is associated with radial filaments, which are oriented in the east-west direction (i.e. perpendicular to the jet axis), located between X-ray cavities, and consistent with radially infalling gas flows. \citet{Lim08} argued that the absence of any large-scale rotation makes a merger-origin of the gas less likely.

The near-infrared spectrum of the nuclear region was investigated by \citet{Rudy93} who found that the [\ion{Fe}{ii}] lines are unusually strong and narrow, indicating that they arise from a region with kinematics different from the nucleus. It was suggested that these features are caused by shocks.
Shocks were also suggested as the cause of the H$_2$ \mbox{1-0} S(1) line emission, first investigated by \citet{Fischer87}. These molecular shocks were discussed as possible triggers of starburst activity. \citet{Krabbe00}, \citet{Rodriguez05}, and \citet{Wilman05} determined the excitation temperatures of the near-infrared H$_2$ lines in the circum-nuclear region. They found the emission of this $\sim 1300-4000$~K H$_2$ component to be consistent with thermal excitation from X-ray or shock heating.
Because the brightest H$_2$ emission is extended over only the central $\sim$300~pc, \citet{Krabbe00} concluded that it is more likely to be directly associated with the AGN rather than being located in a large-scale cooling flow. \citet{Donahue00} agreed with this conclusion and further suggested that the H$_2$ emission is too bright to be part of a cooling flow. In addition, their HST images show a faint extension of H$_2$ \mbox{1-0} S(1) emission from the nucleus to the southwest as well as a faint feature to the east of the nucleus.

\citet{Wilman05} inferred the properties of the near-infrared H$_2$ in the centre of NGC~1275, using observations of the central $6.5 \times 3.4$~arcsec$^2$ with the imaging spectrograph UIST (UKIRT Imaging Spectrometer) on UKIRT (United Kingdom Infrared Telescope) taken under seeing-limited conditions of 0.3--0.4~arcsec. They suggested that the H$_2$ emission at around 50~pc from the nucleus arises in a dense clumpy disc, rotating in a plane perpendicular to the radio jet. They concluded that thermal excitation by the nuclear radiation field is the dominant excitation mechanism of the H$_2$ emission. Based on the discontinuity in H$_2$ velocity across the nucleus of around 240~km~s$^{-1}$, they estimated a mass of $3.4\times10^8$~M$_{\odot}$ for the black hole at the nucleus, assuming a disc inclination of $i=45^\circ$. The nuclear [\ion{Fe}{ii}]~1.644~\micron\ emission was found to be spatially resolved, while the nuclear Pa$\alpha$ emission remained unresolved.
According to \citet{Wilman05}, the nuclear disc of hot H$_2$ would likely be part of the larger-scale H$_2$ and CO structure seen in NGC~1275. This, together with the more recent observations of molecular gas in the extended filaments on several 10~kpc scales \citep{Salome06, Lim12}, suggests that the fuelling process of the AGN in NGC~1275 by interstellar material is being observed over a large range of scales.

In this paper, we analyse the inner circum-nuclear disc of NGC~1275 using observations with NIFS \citep[Near-infrared Integral-Field Spectrograph;][]{McGregor03} in conjunction with the adaptive-optics system, ALTAIR, on the Gemini North telescope. The NIFS data permit a high-spatial-resolution study of the kinematics of the molecular and ionized gas via the H$_2$ \mbox{1-0} S(1), [\ion{Fe}{ii}] 1.644 \micron, \ion{He}{i} 2.058 \micron, and Br$\gamma$~2.166~\micron\ lines as well as an examination of the excitation mechanisms of H$_2$ and [\ion{Fe}{ii}]. 
Throughout the paper we assume H$_0 = 71$~km~s$^{-1}$~Mpc$^{-1}$, $\Omega_M = 0.3$, and $\Omega_{vac} = 0.7$, which results in a spatial scale of 358~pc arcsec$^{-1}$ for NGC~1275 \citep{Wright06}.

\section{OBSERVATIONS}

The Gemini Near-infrared Integral-Field Spectrograph (NIFS) and ALTAIR adaptive-optics system on the 8.1~m Gemini North telescope on Mauna Kea in Hawaii were used to obtain $H$-band and $K$-band spectra of NGC~1275. NIFS has 29 slitlets, each 0.1~arcsec wide, and has a resolution of 0.04~arcsec per spaxel along each slitlet. The field of view is $3.0 \times 3.0$~arcsec$^2$, and it delivers a spectral resolving power of $\sim$ 5300, matched to two pixels on the detector, at both $H$ and $K$. This corresponds to a velocity resolution of $\sim$ 60~km~s$^{-1}$.

The $K$-band observations of NGC~1275 were obtained on 2005 November 11 during the commissioning of NIFS. The observations were performed using the ALTAIR natural guide star system with the nucleus of NGC~1275 as the reference for the adaptive-optics correction. The nearby star U1275\_02199239 was used as the on-instrument wavefront sensor (OIWFS) guide star for slow flexure correction. The instrument was oriented at a position angle of $90^\circ$, so that the slitlets were aligned E-W. The observation consisted of eight exposures of 600~s duration centred on NGC~1275, interleaved with eight 600 s exposures of an adjacent sky region. None of these exposures was spatially dithered. The A0V HIPPARCOS star HIP~18769 was observed after the NGC~1275 observation to provide correction of telluric absorption and flux calibration from its 2MASS magnitudes.

The $H$-band observations of NGC~1275 were obtained during NIFS Science Verification observations (program GN-2006A-SV-124) on 2006 February 1 and 2. Again, ALTAIR was used in natural guide star mode and the instrument was set to a position angle of $90^\circ$. Two object-sky exposure pairs were obtained with 900~s exposure times in modest seeing, one pair on each night. HIP~18769 was also observed after NGC~1275 on each night.

\section{DATA REDUCTION AND VISUALIZATION}
\label{s:reduction}

The data were reduced in a standard way \citep{Storchi-Bergmann09} using a variant of the Gemini IRAF package. The object and sky exposures were combined first. Then a dark exposure was subtracted from both the combined-object and combined-sky frames. The combined-sky frame was scaled to match the combined-object frame and subtracted. The sky-subtracted object frame was then flatfielded, bad pixels were interpolated, and the 29 individual two-dimensional slitlet spectra were extracted from this image. Argon arc spectra were used to define the wavelength calibration and flatfield exposures through a Ronchi grating were used to define the spatial calibration along slitlets. Each two-dimensional slitlet spectrum was then geometrically transformed onto a rectilinear spatio-spectral coordinate grid and the resulting 29 two-dimensional spectra combined into a three-dimensional data cube. A one-dimensional spectrum was extracted from a 0.5~arcsec-diameter circular aperture defined within the standard star data cube, a correction for telluric absorption was defined, and this one-dimensional correction spectrum was applied to all spaxels within the object and standard star data cubes. Similarly, a flux-calibration spectrum was extracted from the telluric-corrected standard star data cube using a 1.5~arcsec-diameter circular aperture with the level set by the 2MASS $H$ or $K$ band magnitude and the blackbody shape defined by its 2MASS $J-K$ colour temperature of 11\,000 K. The final data cubes obtained at $H$ and $K$ are in the form of multi-extension FITS files with 2040 wavelength pixels and $29\times69$ spatial pixels (spaxels).

\begin{figure*}
\centering
\includegraphics[width=\linewidth]{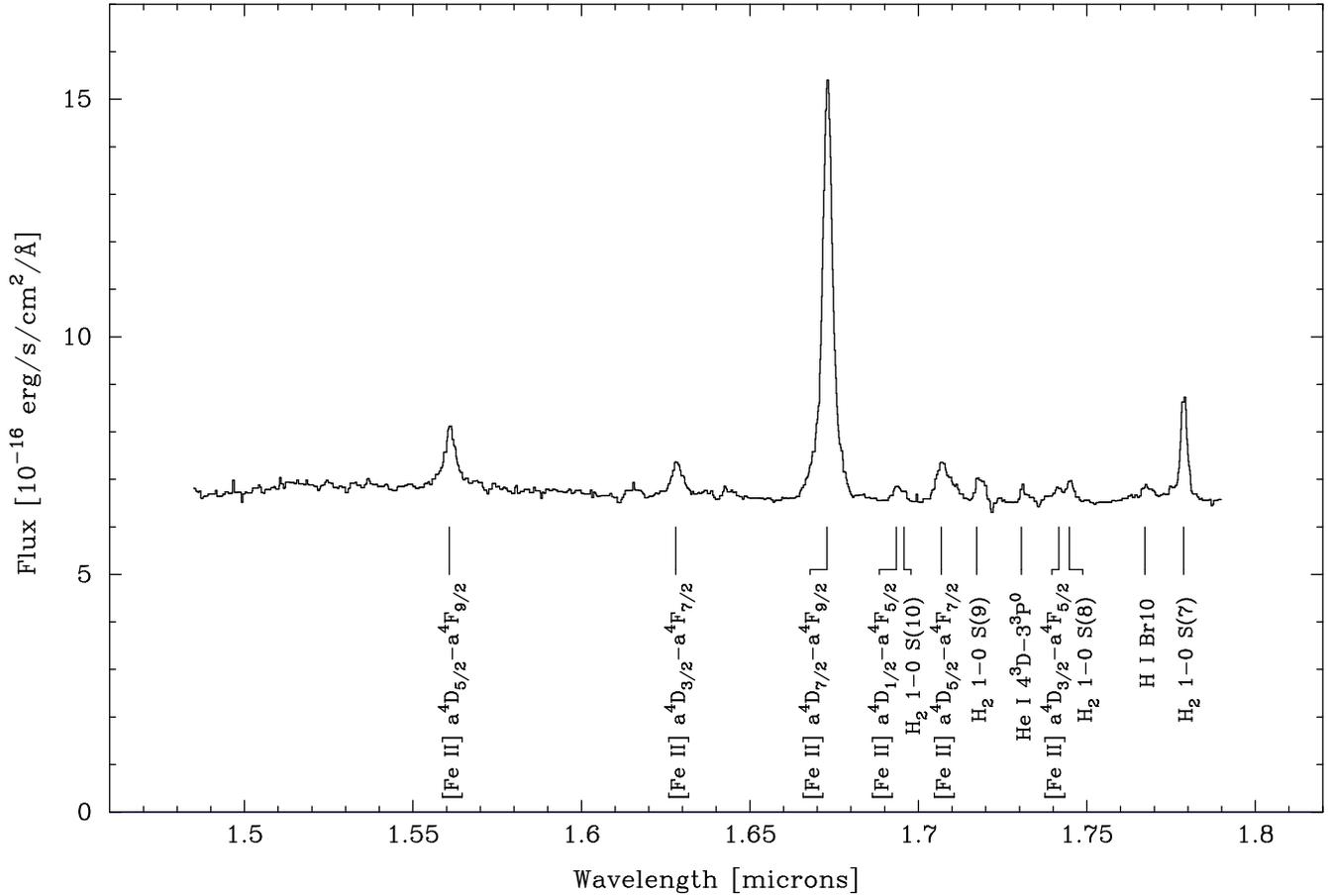}
\caption{$H$-band spectrum of the core of NGC 1275 integrated over a 1.0~arcsec-diameter circular aperture. The wavelength scale is shown as observed for NGC~1275 at $z =  0.017559$. Note that the [\ion{Fe}{ii}] lines are broader than the H$_2$~1-0~S(7) transition.\label{f:HSp}}
\end{figure*}

\begin{figure*}
\centering
\includegraphics[width=\linewidth]{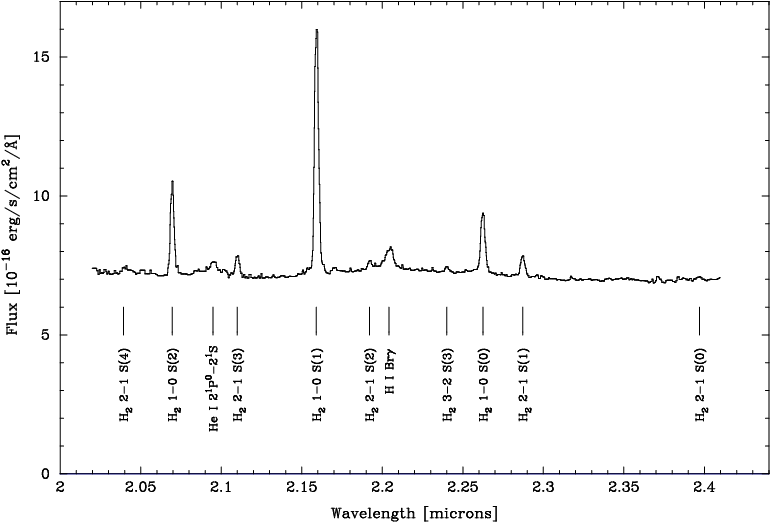}
\caption{$K$-band spectrum of the core of NGC 1275 integrated over a 1.0~arcsec-diameter circular aperture. The wavelength scale is shown as observed for NGC~1275 at $z =  0.017559$. Note the broad base of the \ion{He}{i} and Br$\gamma$ lines, which presumably originate in the inner disc 
(i.e. within the H$_2$ emitting region).\label{f:KSp}}
\end{figure*}

The resulting $H$- and $K$-band spectra of the core of NGC 1275, integrated over a
1.0~arcsec-diameter circular aperture, are shown in Figs~\ref{f:HSp} and \ref{f:KSp}. 
The $K$-band spectrum of the core of NGC~1275 is known to be dominated by H$_2$ emission lines \citep{Krabbe00,Wilman05}.
Our spectra confirm a large number of H$_2$ lines, mostly in the $K$-band spectrum and some weaker lines in the $H$-band spectrum.
The $H$-band is dominated by the strong
[\ion{Fe}{ii}]~1.644~\micron\ emission. Other weaker
[\ion{Fe}{ii}] emission lines are also apparent. 
Furthermore, weaker lines of the Brackett series and of
\ion{He}{i} are seen in the $H$- and $K$-band. The flux measurements for the emission lines detected in the NIFS spectra are listed in Table~\ref{t:Flux}.

\begin{table*}
\centering
\caption{Emission-line fluxes for integrated spectra centred on the nucleus of NGC~1275. Column~1: Name of transition; Column~2: Vacuum wavelength of the line; Columns~3 and 4: Flux and Gaussian FWHM of the lines in the spectrum integrated over a 1~arcsec-diameter
aperture centred on the nucleus; Column~5 and 6: Flux and Gaussian FWHM of the lines in the spectrum integrated over a 0.5~arcsec-diameter
aperture centred on the nucleus. Column~7: Comment regarding unlisted values. The errors of the flux and FWHM measurements are both estimated to be 20~per cent.\label{t:Flux}}
  \begin{tabular}{llccccl}
  \hline
  \multicolumn{1}{c}{(1)}         &               \multicolumn{1}{c}{(2)}                                                & (3)    & (4)      &   (5)   &     (6)  &   \multicolumn{1}{c}{(7)} \\
 \multicolumn{1}{c}{Line}       &     \multicolumn{1}{c}{Vacuum Wavelength}                              & Flux & FWHM & Flux & FWHM & \multicolumn{1}{c}{Comment}\\
                                               & \multicolumn{1}{c}{(\micron)}                                                   & (erg~s$^{-1}$~cm$^{-2}$) &  (km~s$^{-1}$) & (erg~s$^{-1}$~cm$^{-2}$) & (km~s$^{-1}$) & \\
                                             &                                                                                                  & \multicolumn{2}{|c|}{1.0~arcsec aperture} & \multicolumn{2}{c}{0.5~arcsec aperture} & \\
 \hline
$ \mathrm{[\ion{Fe}{ii}]\ a^4D_{5/2}-a^4F_{9/2}}$ & $\qquad 1.533894$ & $5.3 \times 10^{-15}$ & $840$ & $3.2 \times 10^{-15}$ & $900$ &\\
$ \mathrm{[\ion{Fe}{ii}]\ a^4D_{3/2}-a^4F_{7/2}}$ & $\qquad 1.599915$ & $2.7 \times 10^{-15}$ & $740$ & $1.5 \times 10^{-15}$ & $810$ &\\
$\mathrm{[\ion{Fe}{ii}]\ a^4D_{7/2}-a^4F_{9/2}}$ & $\qquad 1.644002$ & $3.7 \times 10^{-14}$ & $690$ & $2.3 \times 10^{-14}$ & $750$ &\\
$ \mathrm{[\ion{Fe}{ii}]\ a^4D_{1/2}-a^4F_{5/2}}$ & $\qquad 1.664225$ & $8.5 \times 10^{-16}$ &  $720$ & ... & ...      & blend\\
H$_2$ 1-0 S(10)                                                   &\qquad  1.6665    & $3.6 \times 10^{-16}$ & ... & $4.0 \times 10^{-16}$ & ... & blend\\
$ \mathrm{[\ion{Fe}{ii}]\ a^4D_{5/2}-a^4F_{7/2}}$ & $\qquad 1.677340$ & $\ 4.4\times10^{-15}$ &  ... & $3.0 \times 10^{-15} $ & ... & blend\\
H$_2$ 1-0 S(9)                                                      & $\qquad 1.6877 $   & $1.5\times10^{-15}$ & $570$ & $6.0 \times 10^{-16}$ & $660$& \\
$\mathrm{\ion{He}{i}\ 4^3D-3^3P^0}$                    & $\qquad 1.70071$  & $6.8\times10^{-16}$ &  ...      & $4.2 \times 10^{-16}$ & ...& blend\\
$ \mathrm{[\ion{Fe}{ii}]\ a^4D_{3/2}-a^4F_{5/2}}$ & $\qquad 1.711599$ & $8.7\times10^{-16}$ & ...      & $4.8 \times 10^{-16}$ & ...& blend\\
H$_2$ 1-0 S(8)                                                      & $\qquad 1.7147  $  & $1.2\times10^{-15}$ &  ...     &$5.6 \times 10^{-16}$ & ...& blend\\
 \ion{H}{i} Br10                                                       & $\qquad 1.7367 $   & $9.1\times10^{-16}$ &   ...    &$5.3 \times 10^{-16}$ & ...& low S/N\\
H$_2$ 1-0 S(7)                                                      & $\qquad 1.7480$    & $5.2\times10^{-15}$ & $390$ & $3.1 \times 10^{-15}$ & $470$& \\
\\
H$_2$ 2-1 S(4)                                                      &$\qquad  2.0041$    & $5.3\times10^{-16}$ &  ... & ... & ...& low S/N \\
H$_2$ 1-0 S(2)                                                      & $\qquad 2.0338$    & $9.2\times10^{-15}$ & $380$ & $5.1 \times 10^{-15}$ & $400$& \\
$ \mathrm{\ion{He}{i}\ 2^1P^0-2^1S}$                   & $\qquad 2.0587$    & $2.1\times10^{-15}$ & $910$ & $1.0 \times 10^{-15}$ & $770$& \\
H$_2$ 2-1 S(3)                                                      & $\qquad 2.0735 $   & $2.2\times10^{-15}$ & $410$ & $1.0 \times 10^{-15}$ & $440$& \\
H$_2$ 1-0 S(1)                                                      & $\qquad 2.1218  $  & $2.9\times10^{-14}$ & $410$ & $1.6 \times 10^{-14}$ & $440$& \\
H$_2$ 2-1 S(2)                                                      & $\qquad 2.1542 $   & $8.1\times10^{-16}$ & $590$ & $4.5 \times 10^{-16}$ & $450$& \\
$ \mathrm{\ion{H}{i}\ Br\gamma}$                         & $\qquad 2.1661 $   & $3.5\times10^{-15}$ & $1000$ & $2.9 \times 10^{-15}$ & $870$&  \\
H$_2$ 3-2 S(3)                                                      & $\qquad 2.2014 $   & $5.6\times10^{-16}$ & $430$ & $3.5 \times 10^{-16}$ & $610$&  \\
H$_2$ 1-0 S(0)                                                      & $\qquad 2.2235 $   & $7.1\times10^{-15}$ & $430$ & $4.2 \times 10^{-15}$ & $500$ & \\
H$_2$ 2-1 S(1)                                                      & $\qquad 2.2477$    & $2.3\times10^{-15}$ & $420$ & $1.1 \times 10^{-15}$ & $460$ & \\
H$_2$ 2-1 S(0)                                                      & $\qquad 2.3556 $   & $2.6\times10^{-16}$ &  ...& ... & ...& low S/N \\
\hline
\end{tabular}
\end{table*}


\begin{figure}
\centering
\includegraphics[width=\linewidth]{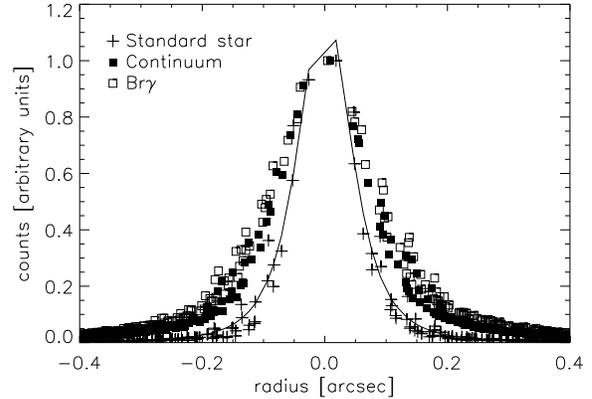}
\caption{PSF fit (solid line) to the $K$-band profile of the standard star HIP~18769 (plus signs) and a comparison with the profiles of the $K$-band continuum (filled squares) and the integrated Br$\gamma$ emission (open squares) of NGC~1275. The Br$\gamma$ emission is integrated over a velocity interval from $-500\ \kms$ to $+500\ \kms$. All profiles are centred on the emission centroids, respectively, and normalized to their peak values.
Positive and negative radii show the profile obtained from an azimuthal average over two opposite halves of the emission peak. The PSF fit is based on a Moffat profile, as explained in the text.\label{f:psf}}
\end{figure}

A knowledge of the $K$-band point-spread function (PSF) is required for analysing the surface-brightness profiles and the kinematics of the emission-line gas in the circum-nuclear region of NGC~1275. We use the spatial profile of the standard star, HIP~18769, as our PSF reference. Obtaining the PSF from separate standard-star observations involves uncertainties, since the observing conditions may have changed between these and the object observations. The PSF could be measured more directly using an unresolved emission component in the $K$-band data cube for NGC~1275. Candidates for this are the $K$-band continuum emission or hydrogen emission from the AGN broad-line region. However, Fig.~\ref{f:psf} shows that the profile of the $K$-band continuum of NGC~1275 (filled squares) is extended compared to the profile of the standard star. The same is true for the Br$\gamma$~2.166~\micron\ emission of NGC~1275 (open squares), which is integrated over a velocity interval from $-500\ \kms$ to $+500\ \kms$ after continuum subtraction. The fact that the Br$\gamma$~2.166~\micron\ emission is spatially extended suggests that the integrated line includes emission from outside the AGN broad-line region. In order to isolate any emission from the AGN broad-line region, it would be necessary to decompose the line profile, which is not practical at the given signal-to-noise ratio.
  
The radial profile of HIP~18769 and the corresponding PSF fit are shown in Fig.~\ref{f:psf}. The PSF fit is based on a single Moffat function of the form
\begin{equation}
P(R) = C_0 \left(1 + \frac{11}{6} \left(\frac{R}{\mathrm{FWHM}}\right)^2\right)^{-11/6}, \label{eq:psf}
\end{equation} 
where $C_0$ is the amplitude, $R$ is the radius, and FWHM is the full width at half maximum. Fitting this function to the standard star profile results in a FWHM of 0.103~arcsec (36~pc), corresponding to one spatial pixel across the slitlets.

\section{EMISSION-LINE MORPHOLOGY}
\label{s:morphology}

\begin{figure}
\centering
\includegraphics[width=\linewidth]{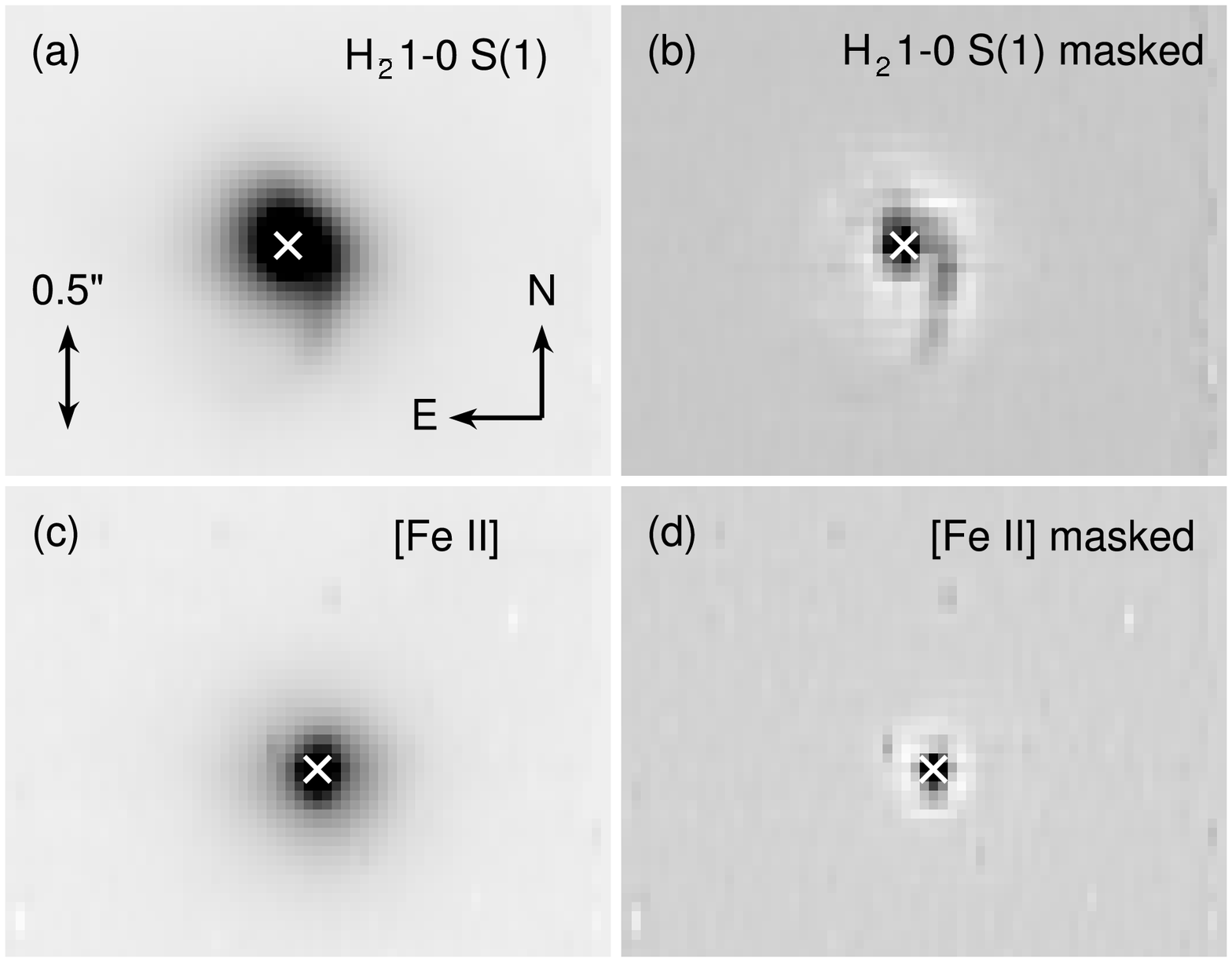}
\caption {(a) Integrated surface brightness of the H$_2$ \mbox{1-0} S(1) emission, (b) unsharp mask of the integrated H$_2$ \mbox{1-0} S(1) surface brightness, (c) integrated surface brightness of the [\ion{Fe}{ii}]~1.644~\micron\ emission, and (d) unsharp mask of the integrated [\ion{Fe}{ii}]~1.644~\micron\ surface brightness. The images are plotted using an inverted grey-scale. The location of the respective continuum peak in the $K$- and $H$-band is indicated with a white cross. The H$_2$ \mbox{1-0} S(1) and [\ion{Fe}{ii}]~1.644~\micron\ emission lines are integrated over velocity intervals from $-500\ \kms$ to $+500\ \kms$ and $-800\ \kms$ to $+800\ \kms$, respectively. The halo of negative pixels around the line emission is an artefact of the unsharp masking procedure, where a smoothed copy of the original image is subtracted from the original image (see text for details). The extended circum-nuclear emission in H$_2$ is interpreted as a large-scale molecular accretion disc (panel a) on which a streamer is superimposed to the south-west of the nucleus (panel b). The streamer is not seen in the [\ion{Fe}{ii}]~1.644~\micron\ emission (panel d).} \label{H2}
\end{figure}
The circum-nuclear regions of the integrated flux maps in H$_2$ \mbox{1-0} S(1) and [\ion{Fe}{ii}]~1.644~\micron\ show a steeply increasing surface brightness towards the core of the galaxy (Fig.~\ref{H2}a and c).
In the analysis presented here, this emission is interpreted as tracing the outer parts of an inclined accretion disc around the central active nucleus. The classical inner accretion disc is unresolved at our spatial resolution of 36~pc.

The integrated H$_2$ emission is clearly perturbed towards the south-west. Fig.~\ref{H2}b shows an unsharp mask of the integrated H$_2$ image of Fig.~\ref{H2}a, which is the result of subtracting a smoothed copy of Fig.~\ref{H2}a from Fig.~\ref{H2}a. The smoothing is performed by convolving the image with a Gaussian function, using a FWHM of 3 pixels. Fig.~\ref{H2}b reveals a feature extending from the nucleus to the west and south, which is likely to be a part of the southwestern emission feature reported by \citet{Donahue00}. It is uncertain whether this is a stream of gas falling into the centre or being ejected from the centre. However, \citet{Lim08} found signs of generally infalling motions in the molecular gas, \mbox{CO(2-1)}, in the inner $R \sim 8$~kpc on kiloparsec scales. Since it is likely that these streams of gas continue to smaller scales, we interpret the southwestern feature in our NIFS data as an infalling stream of gas settling into the overall accretion disc at small radii. As we will see in Sections~\ref{s:kinematics} and \ref{s:model}, the streamer is associated with redshifted velocities and apparently joins the H$_2$ disc on a retrograde orbit. The streamer is not detected in [\ion{Fe}{ii}]~1.644~\micron\ emission. This is obvious from Fig.~\ref{H2}d, which shows the unsharp mask corresponding to the integrated 
 [\ion{Fe}{ii}]~1.644~\micron\ emission. The lack of [\ion{Fe}{ii}] detection in the streamer suggests that the streamer is either predominantly molecular in nature, or not exposed to sufficient internal high-velocity shocks or sufficient X-ray heating from the nuclear continuum source, both of which are possible [\ion{Fe}{ii}] excitation mechanisms.

\section{EMISSION-LINE KINEMATICS}
\label{s:kinematics}

\begin{figure*}
\centering
\includegraphics[width=\linewidth]{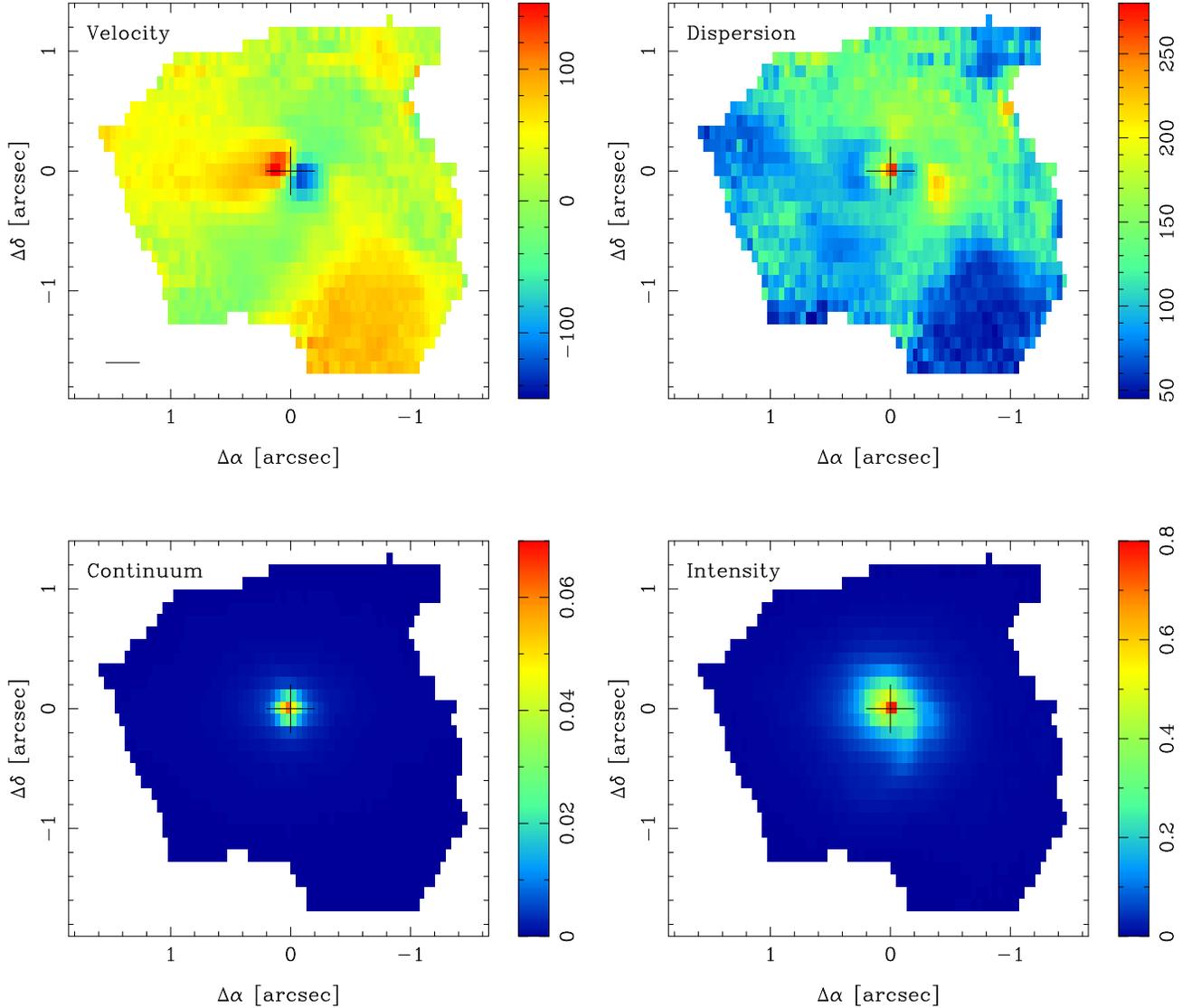}
\caption{Composite result of fitting single-Gaussian profiles to the
H$_2$ \mbox{1-0} S(1)~2.122~\micron\ emission in each NGC~1275 spectrum. The Gaussian central velocity and the Gaussian
dispersion in \kms\ are shown in the upper left and right panels, respectively. The lower-left panel shows the continuum flux density in units of $10^{-15}\ \mathrm{erg\ s^{-1}\ cm^{-2}\ \AA^{-1}}$. The
lower-right panel shows the intensity of the Gaussian fit to the H$_2$ line in units of $10^{-15}\ \mathrm{erg\ s^{-1}\ cm^{-2}}$. The location
of the continuum emission peak is marked by a black cross in each
panel. North is up and east is to the left. The scale bar in the velocity map indicates 100~pc.\label{f:H2Fit}}
\end{figure*}

\begin{figure*}
\centering
\includegraphics[width=\linewidth]{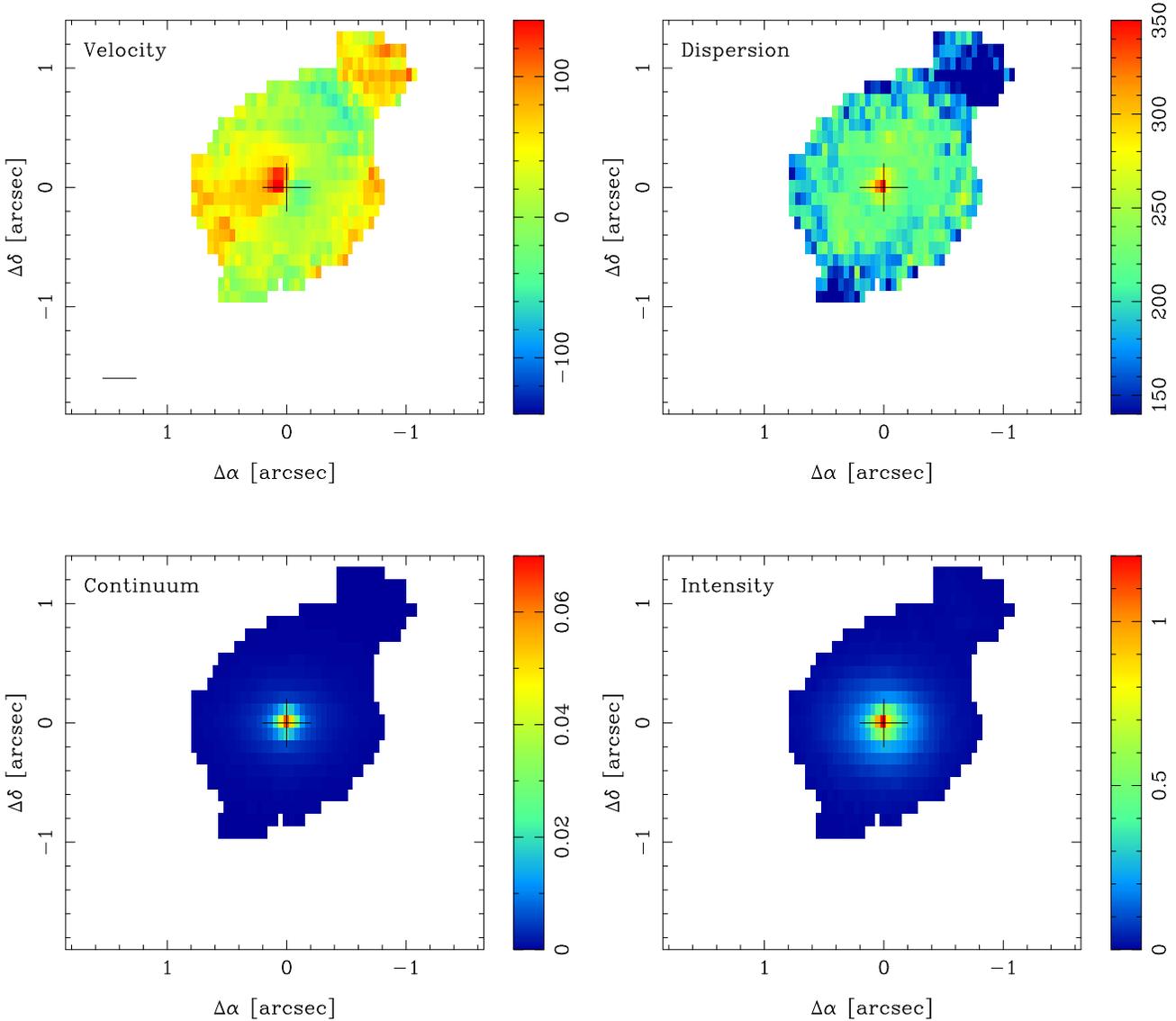}
\caption{Same as Fig.~\ref{f:H2Fit} for the
[\ion{Fe}{ii}]~1.644~\micron\ emission in each NGC~1275 spectrum.\label{f:FeIIFit}}
\end{figure*}

In agreement with the morphological evidence for a circum-nuclear disc from Section~\ref{s:morphology} and the previous indications for a rotating disc based on the lower-resolution data by \citet{Wilman05}, the H$_2$ \mbox{1-0} S(1)~2.122~\micron\ and [\ion{Fe}{ii}]~1.644~\micron\ emission lines show 
kinematics indicative of disc rotation in the
central $R \sim 0.15$~arcsec ($\sim 50$~pc) of NGC~1275.
The kinematic maps, based on a single-Gaussian profile fit to
the H$_2$ \mbox{1-0} S(1)~2.122~\micron\ and  [\ion{Fe}{ii}] 1.644 \micron\ lines, are shown in Figs~\ref{f:H2Fit} and \ref{f:FeIIFit}. 
The four panels of each figure show the central velocity (upper left panel), the velocity dispersion (upper right panel),
the continuum flux density (lower left panel), and the intensity of the Gaussian fit to the line profile (lower right panel).
The scenario of disc rotation is supported by the fact that the redshifted and blueshifted emission of H$_2$ \mbox{1-0} S(1)~2.122~\micron\ and [\ion{Fe}{ii}]~1.644~\micron\ is symmetrical with respect to the peak in continuum intensity (indicated by the cross in Figs.~\ref{f:H2Fit} and \ref{f:FeIIFit}). 
In addition, the continuum peak coincides with the line intensity peak and the peak in velocity dispersion of both lines. 

A jet-driven origin of the H$_2$ and  [\ion{Fe}{ii}] kinematics in the circum-nuclear region can be ruled out, because the major kinematic axis of the proposed disc is perpendicular to the radio jets. The major kinematic axis is found at a position angle of PA=68$^{\circ}$, while the radio jets are located at PA=160$^{\circ}$ \citep{Pedlar90}.
The fact that the two position angles are perpendicular
to each other strongly suggests that the observed rotating disc represents the outer
parts of an accretion disc and that the jets are
ejected along the polar direction. In this case, a disc inclination of about 45$^{\circ}$ can be inferred from the known angle of the radio jets with respect to the line of sight \citep[cf.][]{Wilman05}. Furthermore, since the southern radio jet is, by its brightness, inferred to be pointing towards us, we can conclude that the disc is seen in projection with its northern portions closer to us, and that it is rotating in a counter-clockwise direction.

\begin{figure*}
\centering
\includegraphics[width=\linewidth]{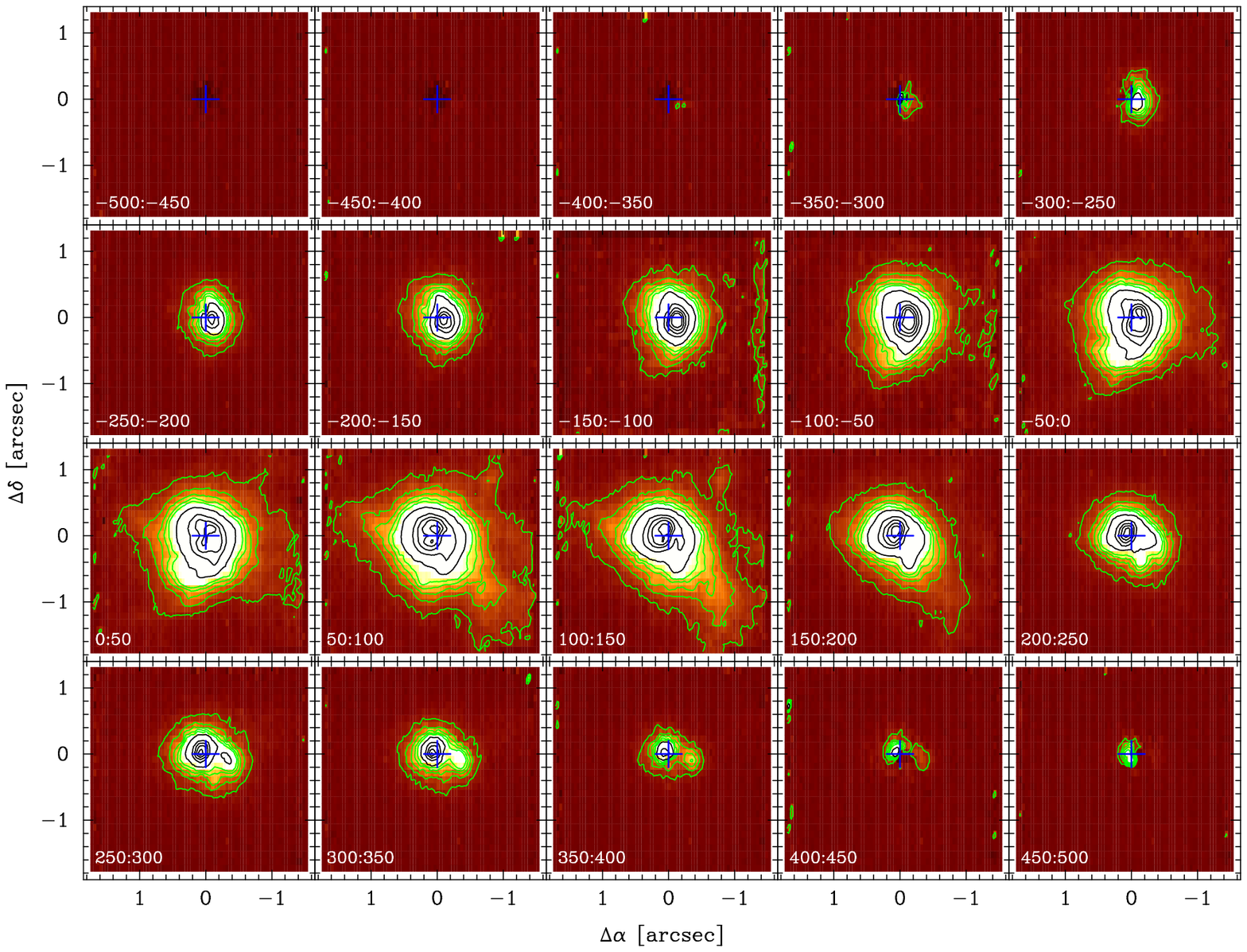}
\caption{Velocity slices through the H$_2$ \mbox{1-0} S(1)~2.122 \micron\
emission from the core of NGC~1275 after subtraction of an average
continuum image. The velocity ranges in \kms\ for each panel are shown
at lower-left. The location of the continuum emission peak is shown by
a blue cross in each panel. North is up and east is to the
left.\label{f:H2Vel}}
\end{figure*}

On larger scales the kinematics of H$_2$ \mbox{1-0} S(1)~2.122~\micron\ and [\ion{Fe}{ii}]~1.644~\micron\ is more complex. Where detected, both lines show a similar line-of-sight velocity field. An exception to this is the region directly south and west of the blueshifted portion of the disc, where the H$_2$ velocity field shows indications of redshifted velocities. The H$_2$ emission from this region is interpreted as a streamer of molecular gas, which is associated with the morphological feature shown in Fig.~\ref{H2}b. The H$_2$ velocity dispersion is characterized by a secondary peak at about 0.4~arcsec west of the nucleus in the upper right panel of Fig.~\ref{f:H2Fit}. At this position the observed H$_2$ line profile splits into separate kinematic components and becomes double-peaked. Since the single-Gaussian fit is not an appropriate model for these complex line profiles, we show the velocity slices for the H$_2$ emission in Fig.~\ref{f:H2Vel}. The dominant streamer feature is seen as an extension to the south-west, which is associated with redshifted velocities. As argued in Section~\ref{s:morphology}, we suggest that this represents an infalling stream of molecular gas. In this scenario, the gas stream enters the nuclear region on a retrograde orbit with respect to the sense of rotation of the molecular disc. This is likely to cause shocks and turbulence that lead to energy dissipation in the disc, which may stimulate more rapid gas accretion onto
the nucleus of NGC 1275. In addition to this prominent streamer, Fig.~\ref{f:H2Vel} shows indications of other streamers which are likely to connect to the filamentary structure around NGC~1275 on larger scales.

A common feature of the kinematic maps in both, H$_2$ \mbox{1-0} S(1) and [\ion{Fe}{ii}]~1.644~\micron\ emission, is the circular region of redshifted velocities 
at a distance of about 1.2~arcsec from the nucleus to the north-west. This region is characterized by small velocity dispersions in both lines. With respect to the core, this region is found at a position angle of $\sim 310^\circ - 340^\circ$, which is close to the projected axis of the radio jet at PA=160$^{\circ}$ \citep{Pedlar90}, (i.e. PA=$340^\circ$ for the northern radio jet with respect to the core). Since the projected location of this region roughly coincides with the jet axis, this feature may result from a jet-cloud interaction.


\section{IONIZATION STRUCTURE}
\label{sec:ionization}

\begin{figure*}
\centering
\includegraphics[width=\linewidth]{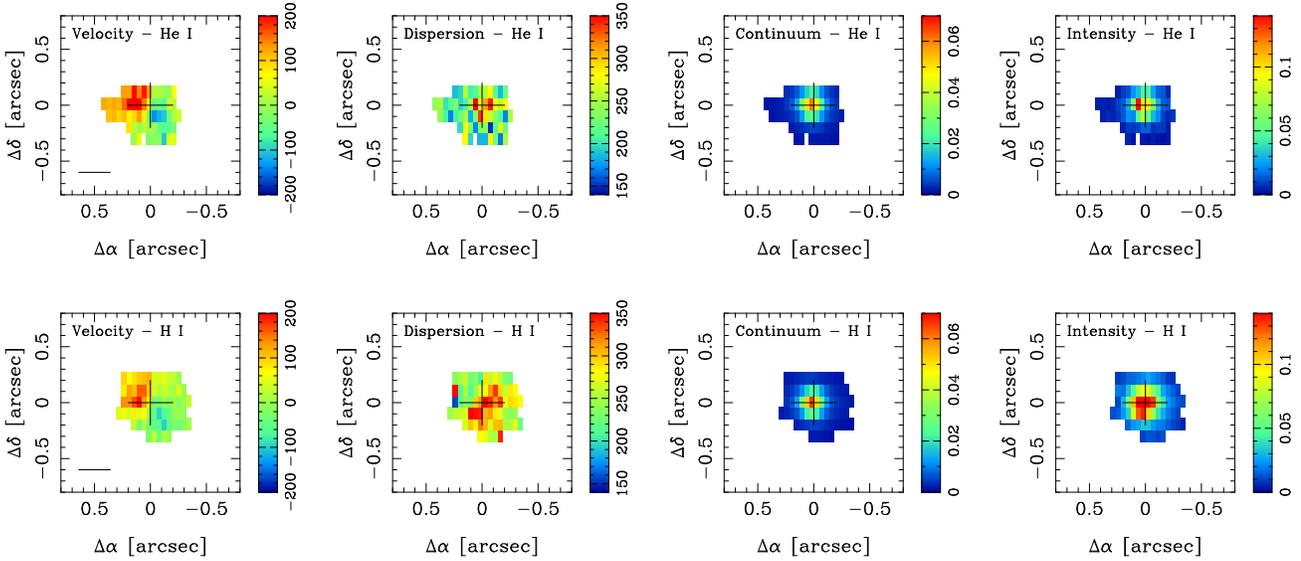}
\caption{Same as Fig.~\ref{f:H2Fit} for the
\ion{He}{i}~2.058~\micron\ emission (top row) and the Br$\gamma$~2.166~\micron\ emission (bottom row) in each NGC~1275 spectrum.\label{f:HeFit}}
\end{figure*}

By comparing the H$_2$ \mbox{1-0} S(1), [\ion{Fe}{ii}]~1.644~\micron\, \ion{He}{i}~2.058~\micron\,  and Br$\gamma$ emission, we find evidence of an ionization structure within the circum-nuclear disc.
H$_2$ \mbox{1-0} S(1) emission traces the outer regions of the disc and is likely to form a toroid, rather than continuing all the way to the nucleus. We use the dust sublimation radius as an estimate for the innermost radius at which H$_2$ molecules can survive without being dissociated by the nuclear radiation field.
A simple estimate of that radius is given by the distance from the core at which the nuclear bolometric luminosity, $L_\mathrm{bol}$, is sufficient for heating the dust to its sublimation temperature, $T_\mathrm{sub}$:
\begin{equation}
L_{\rm bol} = 4\pi R^2_\mathrm{sub} \sigma T_\mathrm{sub}^4 \label{eq:subl}, 
\end{equation}
where $\sigma$ is the Stefan-Boltzmann constant.
We use $4\times 10^{44}\ \mathrm{erg\ s^{-1}}$ as bolometric luminosity of NGC~1275 \citep{Levinson95}. For a dust sublimation temperature of $\sim 1500$~K \citep[see e.g.][]{Nenkova08}, the sublimation radius is about 0.1~pc. A larger sublimation radius is possible, if the sublimation temperature is lower: For example, 1~pc corresponds to a local temperature of about 500~K.
For the forbidden [\ion{Fe}{ii}]~1.644~\micron\ line, the critical density is of the order of $10^5\ \mathrm{cm^{-3}}$ (see Fig.~\ref{f:FeIIdens}). Therefore, the [\ion{Fe}{ii}]~1.644~\micron\ emission becomes less important relative to the permitted lines in the high-density regions closer to the nucleus. The \ion{He}{i}~2.058~\micron\ and Br$\gamma$ lines, however, are expected to be tracers of the (presumably photoionized) inner portions of the disc as well as, possibly, the outer rim of an inflated disc facing the AGN. Evidence for such a scenario can be found in Fig.~\ref{f:HeFit}. This figure shows a single-Gaussian fit to the \ion{He}{i}~2.058~\micron\ and Br$\gamma$ emission lines in NGC~1275, in direct analogy to Figs~\ref{f:H2Fit} and \ref{f:FeIIFit} for the H$_2$ \mbox{1-0} S(1) and [\ion{Fe}{ii}]~1.644~\micron\ emission lines. The \ion{He}{i}~2.058~\micron\ and Br$\gamma$ emission is slightly extended compared to the $K$-band continuum emission of NGC~1275 (see also Fig.~\ref{f:psf}), but more compact than the H$_2$ \mbox{1-0} S(1) and [\ion{Fe}{ii}]~1.644~\micron\ emission. The orientation of the major kinematic axis of the velocity field in \ion{He}{i}~2.058~\micron\ and Br$\gamma$ matches that of the disc in H$_2$ \mbox{1-0} S(1) and [\ion{Fe}{ii}]~1.644~\micron, suggesting that the \ion{He}{i}~2.058~\micron\ and Br$\gamma$ emission are associated with the same physical structure as the latter two emission lines. In addition, like H$_2$ \mbox{1-0} S(1) and [\ion{Fe}{ii}]~1.644~\micron, \ion{He}{i}~2.058~\micron\ and Br$\gamma$ show a peak in velocity dispersion that is centred on the core of NGC~1275. However, the velocity dispersion of \ion{He}{i} and Br$\gamma$ in the core ($\sim 330\ \mathrm{km\ s^{-1}}$) is larger than that of H$_2$ \mbox{1-0} S(1) ($\sim 240\ \mathrm{km\ s^{-1}}$). The same result can be deduced from the integrated nuclear spectrum in Fig.~\ref{f:KSp}, where the \ion{He}{i}~2.058~\micron\ and Br$\gamma$ lines are shown to be broader than the H$_2$ lines (see also FWHM values in Table~\ref{t:Flux}). This larger nuclear velocity dispersion of the \ion{He}{i}~2.058~\micron\ and Br$\gamma$ emission compared to the H$_2$ emission indicates that the \ion{He}{i} and Br$\gamma$ emission arises at least partially from the inner portions of the circum-nuclear disc. These inner parts of the disc are expected to be dominated by high gas velocities that manifest themselves in the form of broad line profiles as a consequence of PSF smearing. Fig.~\ref{f:FeIIFit} shows that the [\ion{Fe}{ii}]~1.644~\micron\ emission in the core of NGC~1275 is also characterized by a larger velocity dispersion -- i.e. a broader line profile in Fig.~\ref{f:HSp} and Table~\ref{t:Flux} -- than the H$_2$ emission. The velocity dispersion displayed by the nuclear [\ion{Fe}{ii}]~1.644~\micron\ emission is comparable to the value found for the \ion{He}{i}~2.058~\micron\ and Br$\gamma$ emission. This suggests that the core of NGC~1275 on scales smaller than the NIFS resolution also contains [\ion{Fe}{ii}]~1.644~\micron-emitting gas with large velocities from ordered or turbulent motion, which may be associated with the disc rotation or with jet-gas interactions.

\section{EXCITATION MECHANISMS}

The excitation mechanism for the H$_2$ and [\ion{Fe}{ii}] emission in the centre of NGC~1275 is analysed based on the 
line flux measurements given in Table~\ref{t:Flux}.

\subsection{Molecular hydrogen}
\label{s:res:h2exc}

\citet{Krabbe00} and, more recently, \citet{Wilman05} and \citet{Rodriguez05}
have analysed
the H$_2$ excitation mechanism in the nucleus of NGC 1275. All find it
to be thermally excited. \citet{Wilman05} find that the excitation
temperature is $\sim 1360 \pm 50$~K from low-excitation transitions,
with a higher temperature of $3100 \pm 500$~K applying to
high-excitation lines, while \citet{Rodriguez05} have used the ratio of the H$_2$ lines to determine a vibration temperature of $2200\pm100$~K, and a rotational temperature of $1300\pm100$~K. 

\begin{figure}
\centering
\includegraphics[width=\linewidth]{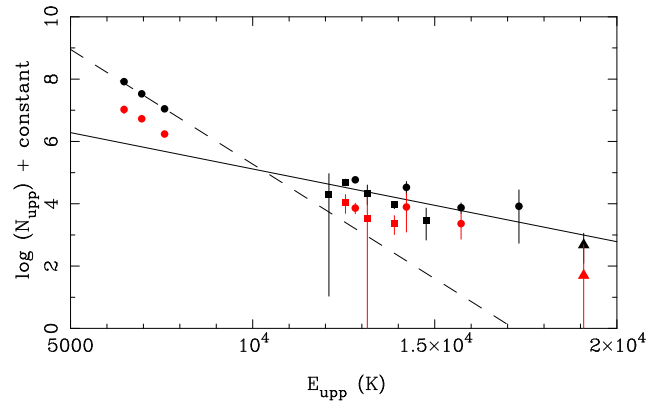}
\caption{H$_2$ population levels in the spectra integrated over a 1.0~arcsec-diameter aperture (black symbols) and a 0.5--1.0~arcsec-diameter annulus (red symbols) centred on the nucleus of
NGC~1275. Relative upper-level population number densities are plotted
versus the upper-level energy, expressed as a temperature. H$_2$~1-0
({\em circles}), 2-1 ({\em squares}), and 3-2 ({\em triangle})
vibrational transitions are shown. Error bars are calculated for a
constant flux error of $\pm 2.5 \times 10^{-16}$~erg~
s$^{-1}$~cm$^{-2}$ for all lines. The dashed straight line corresponds
to the excitation temperature of 1360~K derived by \citet{Wilman05} for
transitions with $E_{\rm up} < 10\,000$~K, normalized here to the H$_2$ 1-0
S(1) transition in the 1.0~arcsec-diameter aperture. The solid straight line shows a least-squares fit to the 
transitions with $E_{\rm up} > 10\,000$~K from the 1.0~arcsec-diameter aperture.\label{f:H2Pop}}
\end{figure}
The relative H$_2$ level populations  for
the central 1.0~arcsec-diameter aperture centred on the nucleus of NGC~1275 are shown in Fig.~\ref{f:H2Pop}, together with the corresponding measurements for a 0.5--1.0~arcsec-diameter annulus. The level populations are
calculated from H$_2$ line fluxes measured from the integrated
$H$-band and $K$-band spectra (Figs~\ref{f:HSp} and \ref{f:KSp}) and listed in Table \ref{t:Flux}. At densities above
$\sim 10^5$~cm$^{-3}$, the H$_2$ gas is expected to be collisionally
excited and so will be in thermal equilibrium. The level populations
are then defined by an excitation temperature, $T_{\rm ex}$, that will be
similar to the local gas temperature and can rise to several thousand
Kelvin through heating by shocks or UV and X-ray irradiation. The
relative level populations are derived by the Boltzman equation such
that $\log(F \lambda / A g) = -E_{\rm up}/T_{\rm ex} + {\rm constant}$,
where $F$ is the line flux, $\lambda$ is the line wavelength, $A$ is
the spontaneous emission coefficient, $g$ is the statistical weight of
the upper level of the transition, and $E_{\rm up}$ is the energy of the
upper level of the transition. These are plotted in the population diagram in Fig.~\ref{f:H2Pop} versus the upper-level energy. Transitions for a gas in
thermal equilibrium lie along a straight line in this diagram with a
slope of $-1/T_{\rm ex}$.

As found by \citet{Krabbe00} and \citet{Wilman05}, H$_2$ transitions from
the core of NGC~1275 with $E_{\rm up} < 10\,000$~K are characterized by an
excitation temperature of $T_{\rm ex} \sim 1400$~K, while transitions from
higher energy levels are characterized by a hotter excitation
temperature. The dashed straight line in Fig.~\ref{f:H2Pop} corresponds to
the $T_{\rm ex} = 1360$~K value found by \citet{Wilman05} for low-energy
transitions. Our data are well fit by this temperature, although our low
energy data are not as extensive as those of \citet{Wilman05} who measured
several Q-branch transitions that lie redwards of our spectra.
A least-squares fit to only higher energy transitions ($E_{\rm up} > 10\,000 \>$K)
gives an excitation temperature of $T_{\rm ex} \sim 4290$~K (solid
straight line in Fig.~\ref{f:H2Pop}). \citet{Wilman05} derive a
comparable value of 3100~K from their data. 

\begin{figure}
\centering
\includegraphics[width=\linewidth]{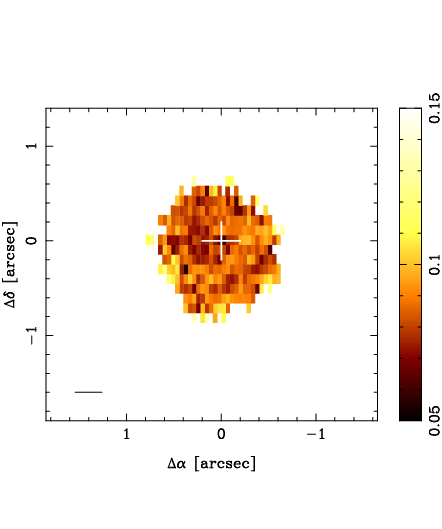}
\caption{Ratio of integrated H$_2$~2-1~S(1)~2.248~\micron\ flux to
H$_2$ \mbox{1-0} S(1)~2.122~\micron\ flux. Each line is integrated over the
velocity range $\pm 250$ \kms. An average continuum has been
subtracted in each case. The location of the continuum emission peak
is shown by a cross. North is up and east is to the
left. The scale bar indicates 100~pc.\label{f:H2Rat}}
\end{figure}

A significant contribution from fluorescent excitation of the H$_2$ emission can be ruled out based on the 
flux ratio of the high-energy H$_2$~2-1~S(1) line
and the low-energy H$_2$ \mbox{1-0} S(1) line, shown in a pixel-by-pixel comparison in Fig.~\ref{f:H2Rat}.
The ratio is spatially uniform at a value of about 0.08-0.09.
Following the diagnostic diagram in fig.~1a of \citet{Mouri94}, this value overlaps with the thermal excitation models but is much lower than the 
typical value of 0.5-0.6 predicted for fluorescent excitation.
Furthermore, the lack of spatial variation in the line ratio over the nuclear region contradicts a scenario with fluorescent excitation. If the
higher excitation temperature of the high-energy levels were due
to irradiation of the gas by nuclear UV or X-ray photons, this effect would then be expected to
be stronger close to the nucleus.
In general, no significant change is
observed in the behaviour of the high-energy transitions in our data
between the 1.0~arcsec-diameter circular aperture and a
0.5--1.0~arcsec-diameter annulus. This can be seen in Fig.~\ref{f:H2Pop}, where the low- and high-energy transitions corresponding to the annulus (red symbols) display the same slope as those corresponding to the circular aperture (black symbols).

\begin{figure}
\centering
\includegraphics[width=\linewidth]{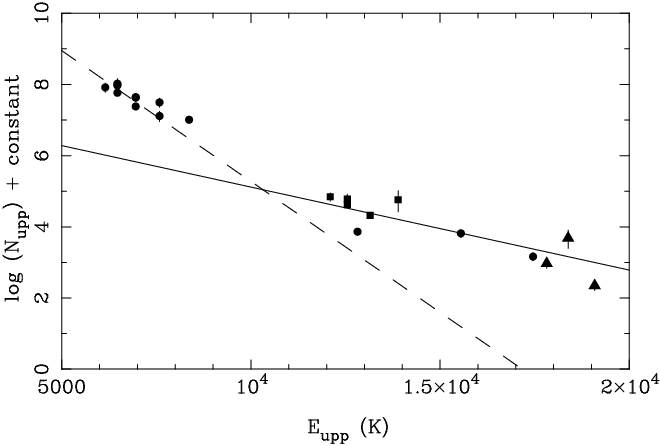}
\caption{Comparison of the population diagram for H$_2$ in the central 1~arcsec of 
NGC~1275 with H$_2$ data for shocked gas in the Orion molecular outflow from 
\citet{Brand88}. The straight lines are reproduced from Fig.~\ref{f:H2Pop}. 
They correspond
to the derived excitation temperatures in NGC~1275 of 1360~K for
transitions with $E_{\rm up} < 10\,000$~K (dashed line) and 4290~K for transitions with $E_{\rm up} > 10\,000$~K (solid line). 
The data points are taken from \citet{Brand88} and represent shocked H$_2$ gas in the Orion molecular outflow. The symbols are the same as in Fig.~\ref{f:H2Pop}.
\label{f:H2PopOrion}}
\end{figure}

The likely scenario for the excitation of the H$_2$ emission in the nuclear region of NGC~1275 is shocks.
In Fig.~\ref{f:H2PopOrion}, we compare the two straight lines corresponding to the excitation temperatures of the low and high-energy transitions in our
H$_2$ data for NGC~1275 to H$_2$ data for shocked gas in the Orion molecular outflow, taken from \citet{Brand88}. The shocked H$_2$ gas in
Orion shows the same properties as our H$_2$ data for NGC~1275 in terms of a higher excitation temperature for higher-energy transitions, presumably arising in post-shock gas having a range of temperatures. A range of excitation temperatures
with higher temperatures for higher-energy transitions is also characteristic of the mid-infrared H$_2$ excitation diagrams
of NGC~1275 and other H$_2$-luminous radio galaxies, which \citet{Ogle10} find to be consistent with shock-excitation.
Therefore, we conclude that shocks are the dominant excitation mechanism for the H$_2$ emission in the central 1.0~arcsec of NGC~1275. 

\subsection{[\ion{Fe}{ii}]}
\label{s:res:feiiexc}

\begin{figure*}
\centering
\includegraphics[angle=-90, width=\linewidth]{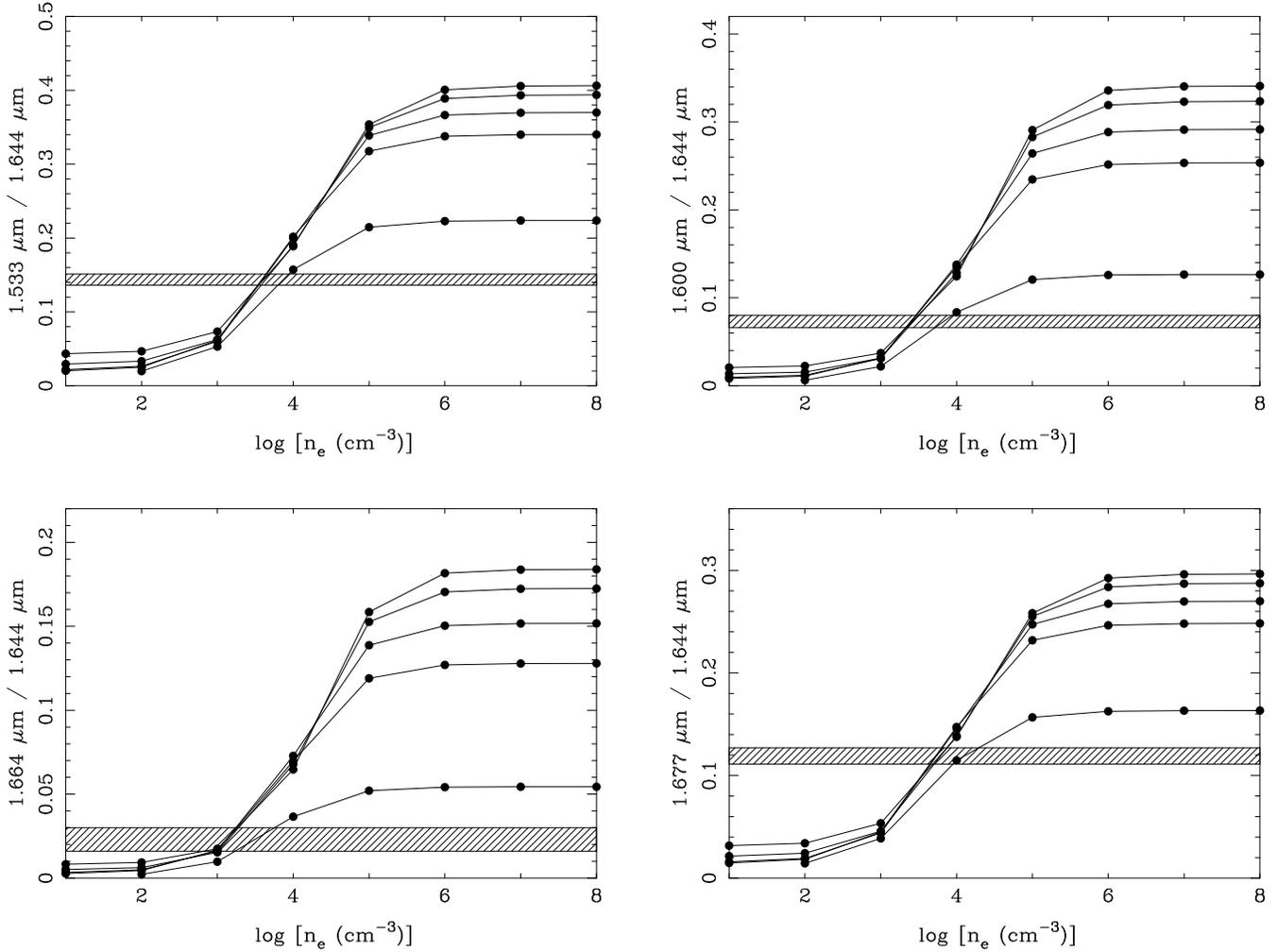}
\caption{Theoretical [\ion{Fe}{ii}] emission-line ratios versus
electron density. The models are computed for a range of electron
temperatures (1, 3, 5, 10, and 20 $\times 10^3 \> {\rm K}$ lowest to
highest in plots). The shaded regions correspond to the measured
line ratios and their uncertainties in the integrated $H$-band spectrum
of NGC 1275 (Fig.\ \ref{f:HSp}).\label{f:FeIIdens}}
\end{figure*}

Several authors have attempted to understand the nature of
[\ion{Fe}{ii}] excitation in NGC 1275 \citep[e.g.][]{Rudy93}. The ionization
potential of Fe$^+$ (16.2 eV) is similar to that of hydrogen, so
[\ion{Fe}{ii}] emission tends to arise in partially ionized
regions. Being due to forbidden transitions between low-energy levels,
the intensities of the near-infrared [\ion{Fe}{ii}] emission lines are
principally dependent on the electron density. Calculations of the
statistical equilibrium of collisional and radiative transitions among
the lowest 16 levels of \ion{Fe}{ii} have been performed by several
authors to study Seyfert galaxies \citep[e.g.,][]{Mouri00, Thompson95} and
young stars \citep[e.g.,][]{Pesenti03, Takami06}. The energy level diagram
for [\ion{Fe}{ii}] has been presented by these authors. Including only
the lowest four terms is justified at the electron temperatures
involved, which are likely to be $< 20\,000\ {\rm K}$ and leave
higher energy levels significantly less populated. We have repeated
this 16-level atom calculation for \ion{Fe}{ii} using level energies
from \citet{Aldenius07}, radiative transition probabilities from
\citet{Quinet96}, \citep[as listed by][]{Aldenius07}, and collision strengths
from \citet{Zhang95} and \citet{Pradhan93}. The model results for a range of
[\ion{Fe}{ii}] emission-line combinations are shown in Fig.~\ref{f:FeIIdens} along with the empirical ratios and their
uncertainties for NGC 1275 measured from our integrated nuclear
spectrum (Fig.~\ref{f:HSp}). The line ratios indicate a density in
the [\ion{Fe}{ii}] region of $\sim 4000 \> {\rm cm^{-3}}$.
\begin{figure}
\centering
\includegraphics[width=\linewidth]{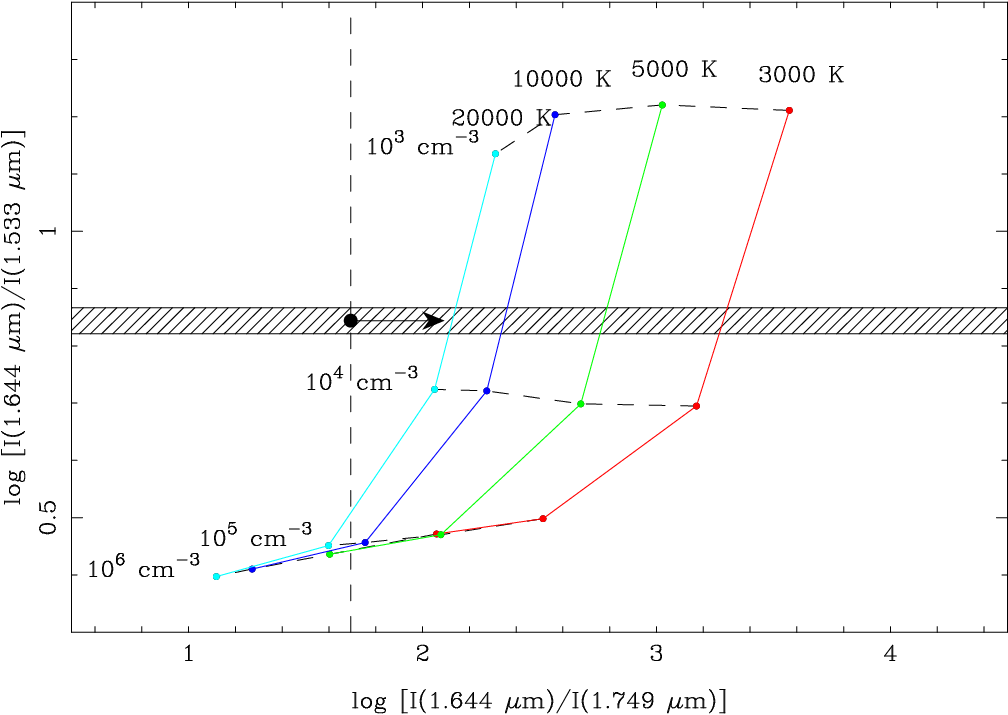}
\caption{Theoretical [\ion{Fe}{ii}] emission-line ratios based on the
a$^4$P$_{3/2}$-a$^4$D$_{7/2}$~1.7489~\micron\ transition. Models are
computed for a range of electron temperatures and electron densities,
as indicated. The shaded region corresponds to the measured 1.644~\micron/1.533~\micron\ ratio and its uncertainty in the
integrated $H$-band spectrum of NGC 1275 (Fig.\ \ref{f:HSp}). The
dashed line and arrow correspond to the limit placed on the 1.644~\micron/1.749~\micron\ ratio by the same data.\label{f:FeII1749}}
\end{figure}
\begin{figure}
\centering
\includegraphics[width=\linewidth]{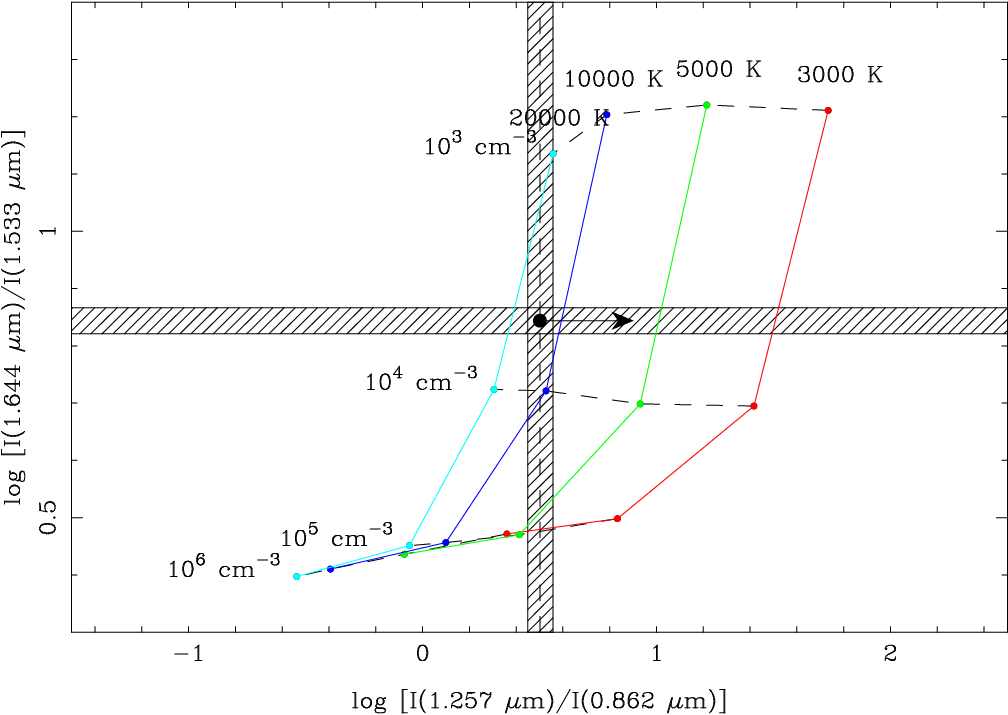}
\caption{Theoretical [\ion{Fe}{ii}] emission-line ratios based on the
a$^4$P$_{5/2}$-a$^4$F$_{9/2}$~0.8617~\micron\ transition. Models are
computed for a range of electron temperatures and electron densities,
as indicated. The horizontal shaded region corresponds to the measured 1.644~\micron/1.533~\micron\ ratio and its uncertainty in the
integrated $H$-band spectrum of NGC 1275 (Fig.\ \ref{f:HSp}). The vertical 
dashed line and arrow correspond to the limit placed on the 1.257~\micron/0.862~\micron\ ratio from \citet{Riffel06}.\label{f:FeII0862}}
\end{figure}

Deriving the electron temperature in the [\ion{Fe}{ii}] region is more
difficult. The near-infrared [\ion{Fe}{ii}] lines arise from
transitions between the lowest three terms (a$^6$D, a$^4$F, and
a$^4$D). The transitions detected in our $H$-band spectra all occur
between the a$^4$D and a$^4$F terms and so have a limited range of
upper-level energies. The $J$-band transitions of [\ion{Fe}{ii}] (e.g.,
at 1.257~\micron) share the same upper energy levels, so do not
provide temperature discrimination. Several authors have suggested
using shorter wavelength transitions from the a$^4$P term or longer
wavelength transitions between the a$^4$F and a$^6$D terms to
constrain the electron temperature in Seyfert galaxies \citep{Mouri00}
and young stars \citep{Nisini02, Pesenti03}. This is the approach
that we adopt here.

We begin by using near-infrared transitions from the a$^4$P energy
level to constrain the electron temperature. The
a$^4$P$_{3/2}$-a$^4$D$_{7/2}$ transition occurs at 1.7489~\micron\ in
the $H$-band so could make a nearly reddening-independent indicator
with the 1.644~\micron\ line. The predicted dependence of this ratio as
a function of electron temperature is shown in Fig.~\ref{f:FeII1749}. In practice, the [\ion{Fe}{ii}]~1.7489~\micron\ line
coincides in wavelength with the H$_2$~1-0~S(7)~1.7480~\micron\ line,
which is strong in NGC 1275. Consequently, we have assumed that the
[\ion{Fe}{ii}]~1.7489~\micron\ line is not detected in our $H$-band
spectrum and have defined a 3$\sigma$ upper limit on the 1.644~\micron/1.749~\micron\ ratio accordingly. This is indicated in
Fig.~\ref{f:FeII1749} by the dashed line and solid arrow. This limit
does not provide a useful constraint on the electron temperature. The
[\ion{Fe}{ii}] a$^4$P$_{3/2}$-a$^4$F$_{3/2}$~0.9474~\micron\ line, which
has been suggested as an electron temperature indicator for young
stars \citep{Nisini02}, is similarly overwhelmed by strong [\ion{S}{iii}]~0.9531~\micron\ emission in NGC 1275 \citep{Riffel06}.

\citet{Mouri00} for Seyfert galaxies and \citet{Pesenti03} for young stars
have both suggested using the [\ion{Fe}{ii}]
a$^4$P$_{5/2}$-a$^4$F$_{9/2}$~0.8619~\micron\ line strength as an indicator
of electron temperature. We choose to compare this line to the  [\ion{Fe}{ii}]~1.257~\micron\
line. This avoids reddening and aperture size issues to the greatest
extent possible. Our model predictions for the dependence of this line
strength on electron temperature are shown in Fig.~\ref{f:FeII0862}. These results are in good agreement with the similar
calculation of \citet{Mouri00}. We use the flux of the [\ion{Fe}{ii}]~1.257~\micron\ line from \citet{Riffel06}.
A prominent emission line is also present at
a rest-frame wavelength of $\sim 8630$~\AA\ in the $I$/$J$-band
spectrum of NGC 1275 presented by \citet{Riffel06}. The feature is at a
significantly different wavelength to the \ion{O}{i}~8446~\AA\ line
seen in several others of their objects and to the closest hydrogen
recombination lines (\ion{H}{i} P13 at 0.8667~\micron\ and P14 at
0.8601~\micron), and also the \ion{Ca}{ii} triplet emission at 0.8498,
0.8542, and 0.8662~\micron. We estimate the line flux to be $\sim
2.0\pm0.2 \times 10^{-14} \>$ erg s$^{-1}$ cm$^{-2}$. If this is a
detection of the [\ion{Fe}{ii}] line, it indicates that the electron
temperature in the [\ion{Fe}{ii}]-emitting region is $\sim 15\,000\ {\rm K}$ (Fig.\ \ref{f:FeII0862}). Otherwise, this represents an upper
limit to the electron temperature. [\ion{Fe}{ii}] 0.8617~\micron\
emission has been seen previously in the Orion Nebula and in the
Seyfert galaxy NGC 4151 \citep{Osterbrock90}, which strengthens the case for
a detection in NGC 1275.

Possible excitation mechanisms for the [\ion{Fe}{ii}] emission
in NGC~1275 are X-ray heating or fast shocks.
\citet{Mouri00} argue that [\ion{Fe}{ii}] excitation by X-ray
heating (i.e., a power-law photoionizing continuum) leads to higher
electron temperatures ($\sim 8000 \> {\rm K}$) than shock heating
($\sim 6000 \> {\rm K}$). To the extent that their models are
applicable, the elevated electron temperature in NGC 1275 is more
indicative of X-ray heating than shock excitation.
The high [\ion{Fe}{ii}]~1.644~\micron/\ion{H}{i}~Br$\gamma$~2.166~\micron\ ratio of $\sim 7.9$ in NGC 1275 also supports X-ray heating as
the dominant excitation mechanism in the nuclear spectrum integrated over a 0.5~arcsec aperture. This ratio is a measure of the relative
volumes of the partially-ionized region emitting [\ion{Fe}{ii}] and the
fully-ionized region emitting \ion{H}{i}. It is expected to be in the
range 0.1--1.4 in starburst regions \citep{Colina93}, but can have
values up to $\sim 20$ in environments where X-rays penetrate deeply
into neutral gas \citep{Alonso97}. The high value of this ratio in NGC
1275 and the proximity of the [\ion{Fe}{ii}] emission to the nucleus
suggest that this is the process that excites the [\ion{Fe}{ii}] emission
in NGC 1275.

We can also use the $J$-band data of \citet{Riffel06} to locate NGC 1275
in the $\epsilon$([\ion{Fe}{ii}]~1.257~\micron)/$\epsilon$([\ion{S}{ii}]~1.03~\micron) versus $\epsilon$([\ion{Fe}{ii}]~1.533~\micron)/$\epsilon$([\ion{Fe}{ii}]~1.644~\micron) 
diagram of \citet{Pesenti03}. They propose this
as a diagnostic of Fe depletion on dust grains in the jets of young
stars. According to table~5 in \citet{Riffel06}, the ratio [\ion{Fe}{ii}]~1.257~\micron/[\ion{S}{ii}]~$(1.029+1.032+1.034+1.037)$~\micron\ in NGC~1275 is 0.5. In contrast, Galactic
Herbig-Haro objects show a ratio of 3.0 \citep{Pesenti03}.
These data, therefore, suggest that Fe is far more
depleted in NGC 1275 than in Galactic
Herbig-Haro objects. This is consistent
with other estimates for Seyfert galaxies \citep[e.g.,][]{Mouri00} where
the strength of the [\ion{Fe}{ii}] emission is interpreted to be
primarily due to the ionization conditions, rather than to significant
dust grain destruction in shocks.

\section{DISCUSSION}

\subsection{Accretion scenario for the molecular disc}
\label{s:accretion}

The NIFS data for the circum-nuclear region of NGC~1275 suggest predominantly shock-excited molecular-hydrogen emission. This emission traces a rotating disc, which, according to
\citet{Wilman05}, is likely to be unstable or to have a clumpy structure, as well as one or several streamer-like features. A possible interpretation of these results is an accretion scenario  
in which gas loses orbital energy as a result of shocks and turbulent dissipation, leading to a net mass inflow and to a net increase in binding energy of the accretion flow. Our model for this scenario is based on the model suggested for the ionized accretion disc in the centre of M87 by \citet[][see their equations (7) and (8)]{Dopita97}. In this model, at each radius in the accretion flow, the net increase in binding energy is matched by the shock luminosity from radiative losses, $L(R)$:
\begin{equation}
L(R)= \frac{d}{dR} \left[ \frac{\dot{M}(R) \phi(R)}{2}\right], \label{shocklum1}
\end{equation}
where $\phi (R)$ is the gravitational potential and $\dot{M}(R)$ is the net mass accretion rate including the inflow rate {\it minus} any mass outflow in a disc wind. 
If the fraction of the total shock luminosity that is radiated into the H$_2$ line is given by $\kappa = S_T/S_\mathrm{H_2}$, then
\begin{equation}
L(R) = 2\pi\,R\,\kappa\,S_\mathrm{H_2}. \label{shocklum2}
\end{equation}

First, we consider the case that the gravitational potential in equation~(\ref{shocklum1}) is dominated entirely by the central massive object ($\phi(R) = \phi_\mathrm{BH}(R) = GM_\mathrm{BH}/R$). In this case, the net gain in binding energy in the accretion flow strongly depends on the radius and increases steeply towards small radii.
Combining equations~(\ref{shocklum1}) and (\ref{shocklum2}), results in
\begin{equation}
\frac{\dot{M}GM_{BH}}{2R^2} = 2\pi\,R\,\kappa\,S_\mathrm{H_2},
\end{equation}
if we assume that the net mass accretion rate $\dot{M}(R)=\dot{M}$ is approximately constant (as is likely in the outer parts of the accretion disc) and also that the ratio $\kappa$ for radiation into the line is approximately constant.
This means that the radial variation in surface brightness should follow a steep scaling law with radius, proportional to $R^{-3}$:
\begin{equation}
S_\mathrm{H_2} \propto \dot{M}M_{BH}{R}^{-3}. \label{SB}
\end{equation}
The assumption of constant accretion rate and radiative efficiency per H$_2$ molecule is clearly not valid for the innermost parts of the accretion disc. In these regions, presumably, disc winds lower the net rate of mass accretion, H$_2$ molecules are dissociated, and the critical density may be exceeded. All these effects lead to a flattening of the H$_2$ surface-brightness profile at small radii. This can be expressed in the form of a core radius so that
\begin{equation}
S_\mathrm{H_2}(R)=\frac{S_0}{1+(R/a_{\rm H_2})^{3}}, \label{eq:H2profile}
\end{equation}
where $S_0$ is the amplitude and $a_\mathrm{H_2}$ is a core radius.

Now, we consider the case that the gravitational potential in equation~(\ref{shocklum1}) is dominated entirely by an extended stellar mass distribution. In this case, the enclosed mass at each radius increases with radius. The corresponding gravitational potential will lead to an accretion-powered H$_2$ surface-brightness profile that is flatter than the one in equation~(\ref{SB}). As the AGN dominates the near-infrared continuum emission of NGC~1275 in the circum-nuclear region, the detailed stellar distribution within the PSF of the AGN is uncertain. However, in order to illustrate the accretion-flow scenario in a gravitational potential that is purely dominated by the stellar mass distribution, we refer to the stellar mass-distribution model used by \citet{Wilman05}. \citet{Wilman05} fitted the stellar surface-brightness profile, obtained from an HST NICMOS F160W image of NGC~1275, at radii $> 0.5$~arcsec, excluding the innermost AGN-dominated radii. They find that the stellar surface-brightness profile can be modelled with a modified Hubble profile with a core radius of $a_\ast =  2.3$~arcsec. For a constant mass-to-light ratio, the modified Hubble profile corresponds to a deprojected stellar density of
\begin{equation}
\rho_\ast (R) = \rho_0 \left[ 1+ \left( \frac{R}{a_\ast}\right)^2 \right]^{-3/2} \label{eq:stelldens}
\end{equation}
\citep[cf.][]{Wilman05}.
Our NIFS observations probe radii which are small compared to the above core radius of $a_\ast =  2.3$~arcsec. On these scales, the stellar density from equation~(\ref{eq:stelldens}) is approximately constant with radius ($\rho_\ast (R) \approx \rho_0$). This means that the enclosed stellar mass increases proportional to $R^3$ and that the gravitational potential $\Phi_\ast (R)$ is proportional to $R^2$. According to equations~(\ref{shocklum1}) and (\ref{shocklum2}), this type of gravitational potential results in an accretion-powered H$_2$ surface-brightness profile that is constant with radius:
\begin{equation}
S_\mathrm{H_2}(R) \approx \mathrm{constant}. \label{eq:H2profilestellar}
\end{equation}

In order to compare the above accretion-flow model for the H$_2$ emission with the data for NGC~1275, we fit the observed H$_2$ surface-brightness profile with a generalized form of equation~(\ref{eq:H2profile}), in which the power-law index $\alpha$ is included as a fit parameter:
\begin{equation}
S_\mathrm{H_2}(R)=\frac{S_0}{1+(R/a_{\rm H_2})^\alpha}. \label{eq:H2fit}
\end{equation}
For $\alpha = 3$, this function is equal to the model for the H$_2$ surface-brightness profile in a gravitational potential that is entirely dominated by the central massive object (equation~(\ref{eq:H2profile})). For $\alpha = 0$, this function is equal to equation~(\ref{eq:H2profilestellar}), which shows the example of an accretion flow in the gravitational potential of an extended stellar mass distribution. In order to take into account PSF smearing, equation~(\ref{eq:H2fit}) is convolved with the PSF model shown in Fig.~\ref{f:psf} before it is fitted to the observed data.
\begin{figure}
\centering
\includegraphics[width=\linewidth]{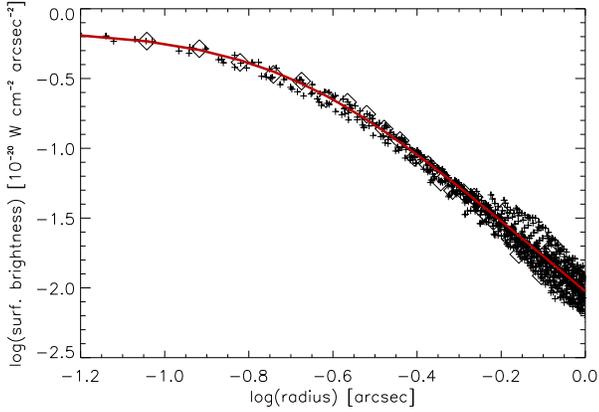}
\caption{Fit to the surface-brightness profile of the integrated H$_2$ \mbox{1-0} S(1) emission of NGC~1275 shown in Fig.~\ref{H2}a. The surface-brightness profile (small plus signs) is obtained over an azimuthal range of $180^{\circ}$, avoiding the perturbed region south-west of the nucleus (see Fig.~\ref{H2}a and b). The large diamonds show the linear interpolation of the profile onto a regular grid in radius. The solid red line shows the fit to the H$_2$ profile using equation~(\ref{eq:H2fit}) convolved with the PSF model shown in Fig.~\ref{f:psf}. The fit results in $\alpha = 2.5$ and $a_{\rm H_2}=0.17$~arcsec.\label{f:H2prof}}
\end{figure}
The observed H$_2$ surface-brightness profile for the inner 1~arcsec (358~pc) of NGC~1275 is displayed in Fig.~\ref{f:H2prof}. The profile (small plus signs) is measured over an azimuthal range of $180^{\circ}$ (in order to avoid the streamer to the south-west of the nucleus) and mapped onto a regular grid in radius via linear interpolation (large diamonds). If the inclination of the H$_2$ disc caused significant projection effects, i.e. if the ellipticity of the isophotes were significant, the measurements for each surface brightness level would be distributed over a range of projected radii. As such projection effects are negligible at $\log(R) < -0.2$~arcsec (225~pc) in Fig.~\ref{f:H2prof} (small crosses), we do not deproject the profile. More significant projection effects are only seen at $\log(R) > -0.2$~arcsec. But the H$_2$ distribution on these scales is likely to be more complex than the disc configuration inferred from the rotation pattern for $R < 50$~pc. 

The fit of the PSF-convolved function of equation~(\ref{eq:H2fit}) to the observed H$_2$ surface-brightness profile results in $a_{\rm H_2}=0.17$~arcsec and $\alpha = 2.5$ (red line in Fig.~\ref{f:H2prof}). This profile is slightly flatter than the profile expected for an accretion flow in a gravitational potential dominated by a central point mass. We, therefore, conclude that the observed H$_2$ surface-brightness profile in NGC~1275 is in general agreement with the above accretion-flow model when taking into account that the gravitational potential has contributions from an extended mass distribution in the inner 1~arcsec (358~pc). 

\citet{Wilman05} estimated that the enclosed stellar mass in NGC~1275 is about 1 per cent of the total dynamical mass at $R=50$~pc (0.14~arcsec). It is therefore negligible on the scales that are characterized by the regular disc rotation (see Section~\ref{s:kinematics}). As the surface-brightness profile is fitted up to larger scales, we use the model of \citet{Wilman05} in order to deduce that the enclosed stellar mass increases by a factor of ten from $R=50$~pc (0.14~arcsec) to $R= 110$~pc (0.3~arcsec) and by a factor of 100 from $R=50$~pc (0.14~arcsec) to $R =240$~pc (0.67~arcsec). 

In addition to the stellar mass, the molecular gas mass is likely to contribute significantly to the total mass in the centre of NGC~1275.
We derive a tentative estimate for the amount of molecular gas in the central $R\sim 50$~pc of NGC~1275 by comparing our NIFS data to the large-scale H$_2$ \mbox{1-0} S(1) and CO observations by \citet{Lim12} and \citet{Salome06}.
\citet{Lim12} report a total H$_2$ \mbox{1-0} S(1) flux of the NGC~1275 nebula (nucleus plus filaments) of $3.87\times 10^{-13}\ \mathrm{erg\ s^{-1}\ cm^{-2}}$. According to our NIFS data, of the order of 1 per cent of this flux originate from the central $R\la 50$~pc. Assuming that this flux ratio translates into a mass ratio and that the same ratio applies to the near-, mid-infrared, and cold molecular gas, we estimate the cold molecular gas mass in the central $R\la 50$~pc to be about 1 per cent of the total cold molecular gas mass of the NGC~1275 nebula \citep[$4\times 10^{10}\ \mathrm{M_\odot}$,][]{Salome06}. This results in a cold molecular gas mass of $M_{\mathrm{H_2}} (R < 50\ \mathrm{pc}) \sim 4\times 10^{8}\ \mathrm{M_\odot}$, which is about two orders of magnitude larger than the stellar mass estimate for the same region.

\subsection{Disc model and black-hole mass}
\label{s:model}

As the H$_2$ emission in the inner $R\sim 50$~pc of NGC~1275 shows evidence for a rotating disc, the enclosed mass of the disc can be inferred from kinematic modelling.
If the potential of the rotating disc is dominated by the central supermassive black hole, the enclosed mass provides a direct estimate of the black-hole mass.
For NGC~1275, a black-hole mass estimate of $3.4\times 10^8$~M$_{\odot} \pm 0.18$~dex was reported by \citet{Wilman05}, using the observed jump in H$_2$ velocity across the nucleus as an indication of disc rotation. Their data, based on seeing-limited UIST (UKIRT Imaging Spectrometer) observations, lack the resolution to reconstruct the full two-dimensional intensity and velocity profiles of the disc. Their black-hole mass estimate is, therefore, based on a single value for the nuclear velocity discontinuity of $240\ \kms$ resolved on a radial scale of 50~pc ($\sim$0.15~arcsec). Another estimate for the black-hole mass of NGC~1275 was given by \citet{Bettoni03}. Using the $M$-$\sigma$ relation and a velocity dispersion of $250\ \kms$, they obtain an indirect mass estimate of $4.1\times 10^8\ \mathrm{M_\odot}$.

We have modelled the H$_2$ emission for the inner $R \sim 50$~pc of NGC~1275 three-dimensionally, using a simulated NIFS data cube 
of a thin, transparent disc of H$_2$ in Keplerian rotation around a central point mass. The model disc is first created in polar coordinates by assigning each radial and azimuthal sampling point a Gaussian line component. The central wavelength of the line component is given by the circular velocity at that radius, as determined from the Keplerian rotation curve. 
The line width is set to the NIFS spectral resolution of $60\ \kms$, i.e. the line components are assumed to be intrinsically unresolved. As discussed in 
Section~\ref{s:accretion}, the observed H$_2$ surface-brightness profile does not show any significant projection effects on $R\sim 50$~pc scales. Therefore, we define the line flux at each radius in the face-on disc using equation~(\ref{eq:H2fit}) with the parameters determined from the fit in Fig.~\ref{f:H2prof}. 
The resulting model disc is projected spatially onto an oversampled Cartesian grid of pixels. This step involves summing up all spectra that fall into each pixel for the given input inclination and the input position angle of PA~$= 68^\circ$ (see Section~\ref{s:kinematics}). Finally, in order to simulate the NIFS data, each two-dimensional slice of the model data cube is convolved spatially, using the PSF from Section~\ref{s:reduction}, and rebinned onto the rectangular NIFS pixel scale. The output is a full $3\times3$~arcsec$^2$ NIFS data cube of an ``observed'' model disc.

For the actual comparison between the model and the data via $\chi^2$ values, we only consider spectral pixels from the central $\sim 0.3 \times 0.3$~arcsec$^2$ ($\sim 110 \times 110\ \mathrm{pc^2}$) around the position of the nucleus of NGC~1275 (on which the model disc is centred). This $\sim 110 \times 110\ \mathrm{pc^2}$ section covers the region with the most regular disc-type velocity field in the observations (see Fig.~\ref{f:H2Fit}). The surface-brightness profile (Fig.~\ref{f:H2prof}) does not show a sharp cut-off at the edges of the kinematically defined disc but rather a smooth decline in H$_2$ surface brightness over the full NIFS field-of-view. Therefore, we use an intrinsic size of the model disc that is much larger than the $\sim 110 \times 110\ \mathrm{pc^2}$ region and extends well beyond this region for all projection angles considered here. For numerical purposes, the model disc is also truncated at small radii in order to avoid infinite velocities when the radius in the Keplerian disc approaches zero. We choose an inner cut-off radius of 1~pc. This corresponds to the minimum radius at which H$_2$ molecules are likely to survive the nuclear radiation field, as outlined in Section~\ref{sec:ionization}.

\begin{figure}
\centering
\includegraphics[width=\linewidth]{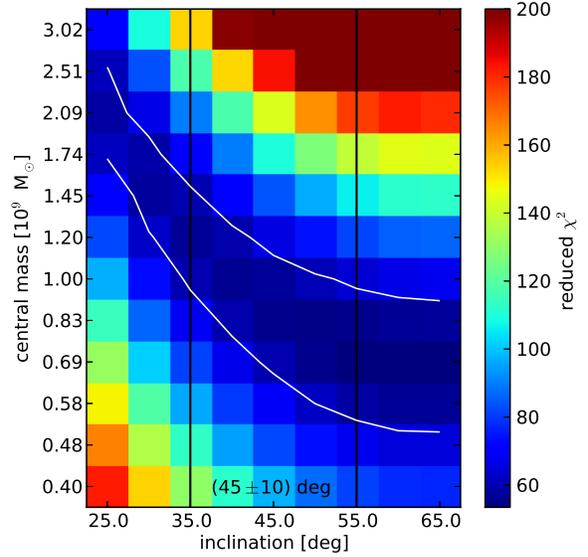}
\caption{Comparison between the disc models and the observed H$_2$ data for a grid of models with varying central mass and inclination.
The reduced $\chi^2$ values are calculated based on a three-dimensional comparison between the modelled and observed data cubes. Spectrally, the $\chi^2$ is computed over a velocity range from $-500 \kms$ to $+500 \kms$ around the rest wavelength of H$_2$ in NGC~1275.
Spatially, the $\chi^2$ values are computed for boxes of $3\times 7$ non-square NIFS pixels (i.e. $\sim 0.3 \times 0.3$~arcsec$^2$ or $\sim 110 \times 110\ \mathrm{pc^2}$) centred on the nucleus, as displayed in Fig.~\ref{f:chi2plot}. The white contour level shows $\Delta \chi^2 = 9$ with respect to the minimum $\chi^2$. The black vertical lines mark the inclination range of $45^\circ \pm 10^\circ$ assumed for the black-hole mass estimate.\label{f:chi2test}}
\end{figure}

\begin{figure*}
\centering
\includegraphics[width=\linewidth]{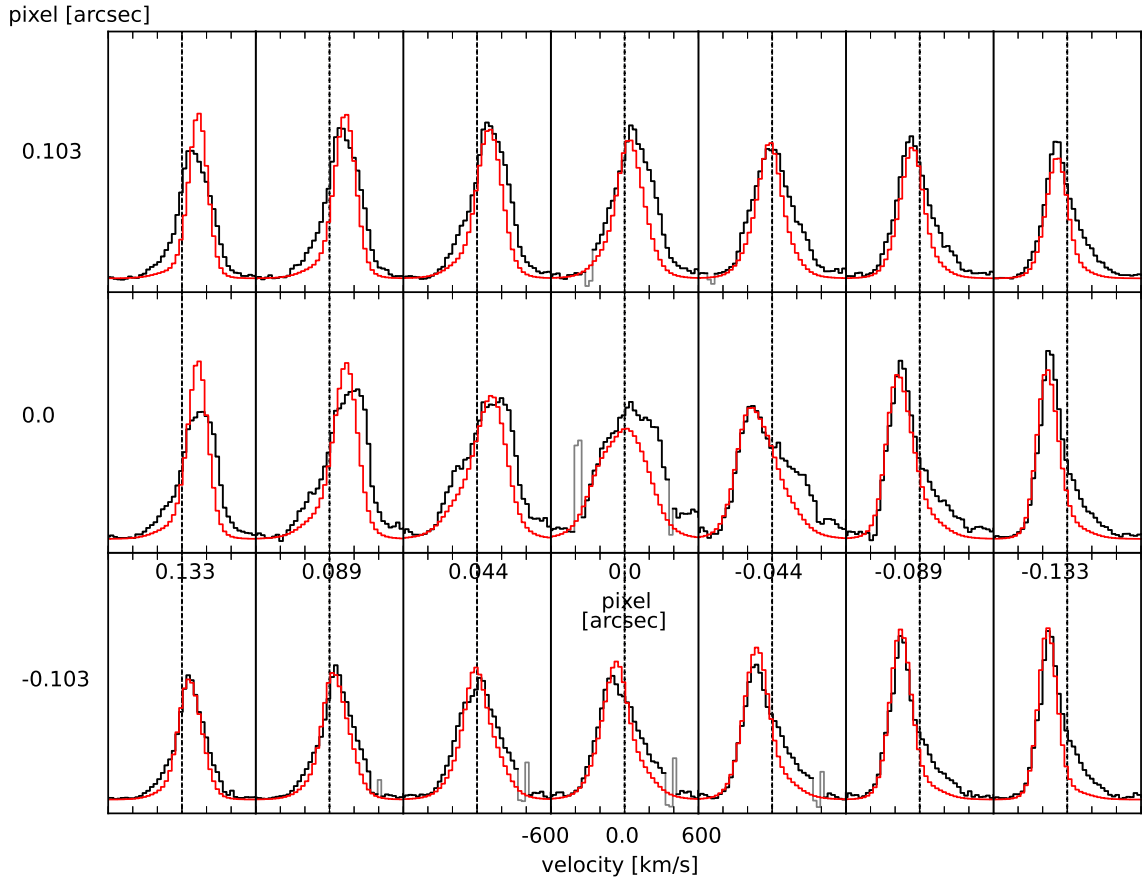}
\caption{Comparison between the H$_2$ profiles in the data (black) and in the disc model for $i=45^\circ$ and $M = 8.32\times 10^8\ \mathrm{M_\odot}$ (red). The figure shows the central $3\times 7$ pixels (i.e. $\sim 0.3 \times 0.3$~arcsec$^2$ or $\sim 110 \times 110\ \mathrm{pc^2}$). In each spectrum, the dotted vertical line indicates the systemic velocity of NGC~1275. Tick marks on the spectral axis are shown in steps of $200\ \kms$. The non-square spatial pixels of NIFS have scales of 0.044~arcsec per pixel in $x$ and 0.103~arcsec per pixel in $y$. Observational artefacts affecting some of the observed spectra are excluded from the $\chi^2$ calculation in Fig.~\ref{f:chi2test} and marked by a light gray colour. North is up, east is to the left.
\label{f:chi2plot}}
\end{figure*}

Fig.~\ref{f:chi2test} shows the $\chi^2$ values for a comparison between the observed H$_2$ \mbox{1-0} S(1) lines and a grid of models with varying central mass and intrinsic disc inclination in the central $\sim 0.3 \times 0.3$~arcsec$^2$ ($\sim 110 \times 110\ \mathrm{pc^2}$) around the core of NGC~1275. In this figure, the absolute flux scale of each model has been adjusted independently to the value that fits the observed H$_2$ lines in the central $\sim 0.3 \times 0.3$~arcsec$^2$ best. As an example, the line profiles from one of the best models ($i=45^\circ$ and $M = 8.32\times 10^8\ \mathrm{M_\odot}$) are compared to the observed line profiles in Fig.~\ref{f:chi2plot}. It is obvious that the observed line profiles in the spectra to the west of the nucleus show a redshifted wing, which is not present in the models for pure disc rotation. This redshifted wing is part of the streamer discussed in Sections~\ref{s:morphology} and \ref{s:kinematics}.

Fig.~\ref{f:chi2test} shows that the enclosed mass of the best-fitting model depends on the disc inclination. This is due to the degeneracy between central mass and inclination in reproducing the same line-of-sight velocities in the Keplerian disc.
Small $\chi^2$ values populate a valley extending from a central mass of about $2.1\times 10^9\ \mathrm{M_\odot}$ at an inclination of $25^\circ$ to a central mass of about $7\times 10^8\ \mathrm{M_\odot}$ at an inclination of $65^\circ$.
The global lower mass limit for the central mass of $\ga 5\times 10^8\ \mathrm{M_\odot}$ in Fig.~\ref{f:chi2test} is robust against a possible underestimate of the PSF width in our $K$-band observations. By increasing the FWHM of the model PSF, the best-fitting models at a fixed inclination are shifted toward higher central masses. This is a consequence of enhanced PSF smearing which causes the central velocity gradient in the models to decrease, so that a higher central mass is required in order to fit the observed velocities. 

\citet{Wilman05} use an inclination of $45^\circ$ in their estimate for the dynamical mass in NGC~1275. This value is based on the assumption that the disc rotation axis has the same orientation as the axis of the radio jet, which is inclined by $30^\circ$-$55^\circ$ with respect to the line-of-sight \citep{Walker94}. For a similar disc inclination of $45^\circ \pm 10^\circ$, our model suggests a best-fitting central mass of $M_{\mathrm{cen}} = (8 ^{\ + 7}_{\ - 2}) \times 10^8\ \mathrm{M_\odot}$ within the contour level shown in Fig.~\ref{f:chi2test}. This is
larger than the $3.4\times 10^8$~M$_{\odot}$ derived by \citet{Wilman05}. This discrepancy is a consequence of mainly two factors: First, the higher resolution NIFS data resolve a larger velocity discontinuity across the nucleus. 
Secondly, in contrast to the calculation by \citet{Wilman05}, our three-dimensional analysis takes into account PSF smearing, which leads to a decrease in
the observed velocities close to the nucleus.
Our value for the central mass is in agreement with the $M$-$\sigma$ relation within the rms scatter. Assuming a stellar velocity dispersion of $\sigma = 250\ \kms$ for NGC~1275 \citep[see][]{Bettoni03}, the updated $M$-$\sigma$ relation for all morphological types from \citet{Graham11}, $\log\ (M_\mathrm{BH}/\mathrm{M_\odot}) = (8.13 \pm 0.05) + (5.13 \pm 0.34) \log\ [\sigma/200\ \mathrm{km\ s^{-1}}]$ results in $M_\mathrm{BH} = 4.2 \times 10^8\ \mathrm{M_\odot}$, which is consistent with our value within the rms scatter of 0.43~dex reported by \citet{Graham11}.

\citet{Wilman05} show that the stellar contribution to the total dynamical mass within $R\sim 50$~pc is negligible and, therefore, argue that the dynamical mass represents the black-hole mass. 
However, as shown in Section~\ref{s:accretion}, we estimate that a large reservoir of molecular gas of the order of $M_{\mathrm{H_2}} (R < 50\ \mathrm{pc}) \sim 4\times 10^{8}\ \mathrm{M_\odot}$ is present in the central parts of NGC~1275. This large gas mass could potentially contribute significantly to the derived enclosed mass. 
In fact, if this gas mass is subtracted from the total enclosed mass of $(8 ^{\ + 7}_{\ - 2}) \times 10^8\ \mathrm{M_\odot}$, the resulting black-hole mass of $\sim 4 \times 10^8\ \mathrm{M_\odot}$ is almost exactly the same as the one derived from the $M$-$\sigma$ relation. However, as this result involves large uncertainties, we conclude by merely noting that our estimate of the enclosed mass may rather represent an upper limit for the black-hole mass.

\subsection{Comparison to other H$_2$-luminous radio galaxies}
\label{s:comparison}

Using {\it Spitzer} observations of pure rotational H$_2$ transitions in the mid-infrared, \citet{Ogle10} detect warm molecular hydrogen emission in 30 per cent of their sample of $z < 0.2$ 3C radio galaxies, including NGC~1275. These molecular hydrogen emission galaxies (MOHEGs) are found among radio galaxies of both Fanaroff--Riley classes and among cool core clusters. The sample studied by \citet{Ogle10} shows that radio MOHEGs as a class typically have low star-formation rates despite the large H$_2$ masses, which indicates that turbulence prevents the gas from efficiently forming stars. Furthermore, most MOHEGs harbour low-luminosity AGN that cannot provide sufficient X-ray heating for powering the observed H$_2$ emission. 
The AGN in NGC~1275 ($\log L_X (2-10\ \mathrm{keV}) = 42.91$) belongs to the more X-ray luminous AGN among the radio MOHEGs \citep[cf. fig.~17 and table~9 in][]{Ogle10}. This means that X-ray heating may make a stronger contribution in NGC~1275 than in many X-ray faint MOHEGs. However, fig.~17 in \citet{Ogle10} shows that the ratio of the observed warm H$_2$ luminosity to X-ray luminosity for NGC~1275, i.e. $L(\mathrm{H_2})/L_X =  0.07$, is still a factor of 7 higher than the maximum ratio of 0.01 derived for an X-ray dissociation region. Therefore, even in NGC~1275, X-ray heating by the AGN is unlikely to be the primary excitation mechanism for H$_2$.

Instead, \citet{Ogle10} conclude that the most likely excitation mechanism for the H$_2$ emission in radio MOHEGs is shocks (or, less likely, cosmic ray heating).
The same mechanism is suggested for NGC~1275, based on the NIFS data discussed in this paper (Section~\ref{s:res:h2exc}). This raises the question whether these shocks originate from radio-jet feedback, as suggested for the overall sample of radio MOHEGs by \citet{Ogle10}, or from gas accretion, as discussed for the molecular disc in NGC~1275 in Section~\ref{s:accretion}. In the feedback scenario, shocks are driven into dense clouds in a porous interstellar medium by the hot cocoon powered by the radio jet, as for example shown in the simulations by \citet{Wagner11} and \citet{Sutherland07}. In the accretion scenario, gas clumps are repeatedly shocked while falling into the centre of the galaxy so that gravitational potential energy is converted into shocks and turbulent heating. 
In a case study of 3C~326~N, a $z=0.09$ radio MOHEG with a very low star-formation rate and an X-ray faint AGN, \citet{Ogle07} suggest a tidally-induced accretion flow as the most likely origin of the shocks, while \citet{Nesvadba10} favour the jet feedback scenario. 

Simple energy considerations for NGC~1275 show that either of the processes, radio jet feedback or gas accretion, is sufficient to power the observed mid- or near-infrared H$_2$ emission. 
For the radio jet, \citet{Ogle10} report a jet cavity power of $P_{\mathrm{Jet\ Cavity}} = 1.4\times10^{44}\ \mathrm{erg\ s^{-1}}$, which is almost 3 orders of magnitude larger than the luminosity of the warm H$_2$ emission \citep[$10^{41.75}\ \mathrm{erg\ s^{-1}}$, cf.][]{Ogle10}. 
For the accretion process, we estimate an accretion luminosity of the order of $10^{43}\ \mathrm{erg\ s^{-1}}$, which is about one order of magnitude larger than the warm H$_2$ luminosity.
This accretion luminosity is derived by dividing the potential energy of the infalling gas of about $3\times 10^{58}\ \mathrm{erg}$ by the dynamical free-fall time of $\sim 1\times 10^8$~yrs, in analogy to \citet{Ogle07}. For this calculation we assume that the total cold gas mass of $4\times 10^{10}\ \mathrm{M_\odot}$ \citep{Salome06} falls back from an average distance of 30~kpc into the potential well corresponding to a stellar mass of $\sim 3\times 10^{11}\ \mathrm{M_\odot}$. This stellar mass is derived from the $M_{\mathrm{BH}}$-$M_{\mathrm{bulge/host}}$ relations at $z=0$ by \citep{Bennert11}, using the $M_\mathrm{cen}$ from Section~\ref{s:model} as the black-hole mass. 

As neither jet feedback nor gas accretion stands out as the dominant energy source in NGC~1275, it is likely that the observed properties of gas excitation and kinematics are a consequence of both processes.
Rotating molecular discs, as found in the circum-nuclear region of NGC~1275, or on scales of 3~kpc in 3C~326~N \citep{Nesvadba11}, suggest that accretion processes play a role.
Since the radio jet cannot impart angular momentum on a gaseous disc, disc rotation is a sign of gas infall. This gas inflow may be associated with tidal galaxy-galaxy interactions, or with accretion from the halo. These processes have been discussed for the case of 3C~326~N \citep{Ogle07, Nesvadba10}. For NGC~1275, it is most likely that the gas inflows consist of gas which falls back from the radio jet bubbles \citep{Salome08}. 
Such back flows of gas uplifted during the active jet phase are also seen in simulations \citep{Sutherland07}. Tidally-induced gas inflows could result from an interaction with the high-velocity system. But this scenario is less likely, since the high-velocity system is still at a large distance from NGC~1275 \citep{Gillmon04}.  

Observational evidence suggests that turbulence is a common characteristic of the molecular gas component in radio MOHEGs.
\citet{Nesvadba11} found the molecular gas in the disc of 3C~326~N to be highly turbulent with line widths of FWHM~$\sim 500\ \kms$. Broad H$_2$ lines are also characteristic of the sample of radio galaxies with fast ionized/atomic outflows, studied by \citet{Guillard12}. Integrated over the full field-of-view, our NIFS data of NGC~1275 show an H$_2$ \mbox{1-0} S(1) line width of FWHM~$\sim 350\ \kms$, which is comparable to the above values. A fraction of this line width is likely to be attributed to ordered streaming motions, which are characterized by small ($\la 100\ \kms$) and largely positive velocities (see Fig.~\ref{f:H2Vel}). Since the models for the circum-nuclear disc in NGC~1275 (Section~\ref{s:model}) provide a good representation of the data without including any significant intrinsic line width, the velocity dispersion of the H$_2$ \mbox{1-0} S(1) lines are likely to be $\la 100\ \kms$ (i.e. FWHM~$\la 240\ \kms$). (The total widths of the H$_2$ lines in the small-aperture spectra, listed in Table~\ref{t:Flux}, are much larger than that, but are dominated by the disc rotation.) The fact that the dispersion-related line width in the disc is smaller than the total integrated line width in the NIFS field-of-view suggests that the accretion flow is energetically capable of producing the turbulent motions in the disc. Alternatively, the turbulence may be driven by the radio jet and its hot cocoon. Turbulent heating via radio jet feedback can affect a large volume if the gas is distributed in the form of clumps in a porous interstellar medium \citep{Wagner11}.

Turbulent heating of the molecular gas component by the jet interaction is only compatible with the observed disc configurations if the jet is, at the same time, inefficient in expelling the molecular gas from the disc.
Such a picture is also suggested by the observations by \citet{Guillard12}, who find that in radio galaxies with fast ionized and atomic outflows the bulk of the mid-infrared H$_2$ emission does not participate in the outflow, despite showing signs of turbulent heating. Simulations by \citet{Sutherland07}, which probe the effects of a jet interaction with an almost Keplerian and turbulently supported disc of a warm clumpy interstellar medium in an elliptical galaxy, show that the jet interaction is, indeed, inefficient in driving the gas out of the disc region. However, there is a significant gas transport from the inner to the outer portions of the disc, which will eventually leave the inner portions depleted. If the disc is porous, the mass transfer rates are lower than in the case of a homogeneous disc of gas. 

We estimate the porosity of the circum-nuclear disc in NGC~1275 by comparing the observed mean density of H$_2$ molecules in the central $R\sim 50$~pc with the density expected for the corresponding gas component from the ideal equation of state. As shown in Section~\ref{s:accretion}, we tentatively estimate the total molecular gas mass within a radius of $R\sim 50$~pc to be $M_{\mathrm{H_2}} (R < 50\ \mathrm{pc}) \sim 4\times 10^{8}\ \mathrm{M_\odot}$. The corresponding mean number density of molecules in the central volume is $\bar{n}_{\mathrm{H_2}} = M_{\mathrm{H_2}} (R < 50\ \mathrm{pc}) /(2 m_p V) \sim 2\times 10^4\ \mathrm{cm^{-3}}$, where $m_p$ is the proton mass and $V$ is approximated by the spherical volume enclosed by a radius of 50~pc. This value is similar to the clump density derived from the ideal equation of state, if we assume that the clumps are composed of cold molecular gas at a mean temperature of $T=50$~K and that they are in pressure equilibrium with the hot ISM pressure of $p/k = 10^6\ \mathrm{cm^{-3}\ K}$, following the scenario by \citet{Wagner11}. The similarity between the mean observed density and the estimated clump density from the ideal equation of state suggests that the cold molecular gas, if distributed in clumps, is likely to have a high filling factor. An independent measure of the actual gas density is provided by detection experiments for high-density gas tracers. \citet{Salome08b} report the detection of HCN(3-2) in the centre of NGC~1275, which indicates a gas density of at least $10^4\ \mathrm{cm^{-3}}$. This value agrees with the mean gas density estimated above. But, as a lower limit for the actual gas density, it does not rule out the possibility of smaller filling factors. Detections of other high-density gas tracers are needed in order to obtain a better estimate of the filling factor.

The relative importance of accretion and jet feedback in exciting and heating the molecular gas in NGC~1275 and other radio MOHEGs is poorly constrained with the current data. Improved observational parameters for the molecular gas phase  -- e.g., from ALMA observations of high-density tracers -- together with refined simulations are required in order to explore the effects of the jet interaction with the interstellar medium in these molecular-hydrogen-rich radio galaxies.

\section{SUMMARY AND CONCLUSIONS}

The paper is based on near-infrared integral field spectroscopy of NGC~1275, obtained with the NIFS instrument at Gemini North at a spatial resolution of $\sim 0.1$~arcsec. We have presented a scenario in which the shock-excited molecular hydrogen, traced via the near-infrared ro-vibrational H$_2$ lines, enters into the circum-nuclear region in the form of one or more streamers and settles into a turbulent, clumpy or unstable accretion disc close to the nucleus, where energy dissipation via turbulence and shocks drives the further accretion process towards the core. The observed disc kinematics has been used as a tool to infer a direct estimate for the central mass enclosed by the disc via model data cubes of discs in Keplerian rotation. We have compared the results to other H$_2$-luminous radio galaxies and discussed the relative importance of jet feedback and accretion as a mechanism for shock-excitation and heating of the molecular hydrogen.

In summary, the main results presented in this paper are:
\begin{itemize}
\item The circum-nuclear H$_2$ \mbox{1-0} S(1) and  [\ion{Fe}{ii}]~1.644~\micron\ lines in NGC~1275 show evidence for a disc on $R=50$~pc scales, both in morphology and kinematics. The projected major kinematic axis of this disc is perpendicular to the projected axis of the radio jets. 
\item A possible ionization structure within the disc is discussed by comparing the H$_2$ \mbox{1-0} S(1) and [\ion{Fe}{ii}]~1.644~\micron\ emission with the \ion{He}{i}~2.058~\micron\ and Br$\gamma$ lines. The latter two lines are only found in a compact region around the core, but their kinematics is characterized by the same disc signature as seen in H$_2$ \mbox{1-0} S(1) and  [\ion{Fe}{ii}]~1.644~\micron. It is suggested that both the \ion{He}{i}~2.058~\micron\ and Br$\gamma$ emission lines arise at least partially from the inner parts of the disc, because they display a larger central velocity dispersion than the H$_2$ \mbox{1-0} S(1) line. Future higher angular resolution observations are required to fully explore this ionization structure.
\item The H$_2$ surface-brightness profile in the inner 1~arcsec around the core is well represented by a PSF-convolved profile following an $R^{-2.5}$ power law with an inner core radius of 0.17~arcsec. This type of surface-brightness profile is found to be in general agreement with an accretion model in which the line emission matches the net increase in binding energy of gas falling into the combined gravitational potential of a central massive object and an extended mass distribution.
\item In H$_2$, the data reveal an elongated perturbation to the south-west of the nucleus, which is characterized by redshifted velocities. This perturbation is interpreted as the strongest of several molecular gas streamers, which are likely to be falling into the core.
\item The relative line strengths of the ro-vibrational H$_2$ lines in the central 1~arcsec of NGC~1275 indicate thermal excitation of the molecular hydrogen via shocks. In agreement with previous studies, we find excitation temperatures of 1360~K and 4290~K for transitions with $E_{upp} < 10\,000$~K and $E_{upp} > 10\,000$~K, respectively. Any significant contribution from fluorescent excitation can be excluded based on the small ratio of the H$_2$~2-1~S(1)~2.248~\micron\ to the 
H$_2$ \mbox{1-0} S(1)~2.122~\micron\ flux and its spatial uniformity over the nuclear region. Instead, the H$_2$ excitation in NGC~1275 shows the same characteristics as shocked gas in the Orion molecular outflow. 
\item The high electron temperature as well as the high [\ion{Fe}{ii}]~1.644~\micron/\ion{H}{i}~Br$\gamma$~2.166~\micron\ suggest that the [\ion{Fe}{ii}] lines in the integrated nuclear spectrum are excited by X-ray heating. Alternatively, the line emission can be excited by fast shocks. Models for a range of [\ion{Fe}{ii}] emission-line ratios are presented. They indicate a density of $\sim 4000\ \mathrm{cm^{-3}}$ in the [\ion{Fe}{ii}] emitting region. The electron temperature in this region is less well constraint. Using less certain detections of [\ion{Fe}{ii}] lines at shorter wavelengths from previous publications, an electron temperature of $\sim 15\, 000$~K is estimated. 
\item An estimate for the mass enclosed by the circum-nuclear H$_2$ disc of NGC~1275 is derived by modelling the data via simulated NIFS data cubes. Assuming a disc inclination of $45^\circ \pm 10^\circ$, the enclosed mass is estimated to be $(8 ^{\ +7}_{\ - 2}) \times 10^8\ \mathrm{M_\odot}$. Interpreted as a direct measurement of the black-hole mass, this value is larger than previous black-hole mass estimates, but is consistent with the $M$-$\sigma$ relation within the rms scatter. A tentative estimate of the molecular gas mass in the central region of NGC~1275 results in $M_{\mathrm{H_2}} (R < 50\ \mathrm{pc}) \sim 4\times 10^{8}\ \mathrm{M_\odot}$, which is a non-negligible fraction of the enclosed mass. The enclosed mass may, therefore, represent an estimate for the upper limit of the black-hole mass.
\item Basic energy considerations show that the shock-excitation and turbulent heating of the molecular gas in NGC~1275 could be driven by the jet interaction, the accretion process, or a mixture of both. An order-of-magnitude estimate of the distribution of the cold molecular gas in the circum-nuclear disc of NGC~1275 indicates a high filling factor. Improved constraints on the properties of the molecular gas phase can be expected from future observations (e.g., ALMA).
\end{itemize}

\section*{ACKNOWLEDGMENTS}
We are grateful to the anonymous referee for very useful comments and suggestions. We thank Alex Wagner for a helpful discussion.
We acknowledge the tireless efforts of the NIFS team from The Australian National University, Auspace, and the Gemini Observatory who made the NIFS instrument reality. This work is based on observations obtained at the Gemini Observatory, which is operated by the Association of Universities for Research in Astronomy, Inc., under a cooperative agreement with the NSF on behalf of the Gemini partnership: the National Science Foundation (United States), the Science and Technology Facilities Council (United Kingdom), the National Research Council (Canada), CONICYT (Chile), the Australian Research Council (Australia), Minist\'erio da Ci\^encia, Tecnologia e Inova\c{c}\~{a}o (Brazil), and the Ministerio de Ciencia, Tecnolog\'ia e Innovaci\'on Productiva (Argentina).
J. S., P. J. M., and M. A. D. acknowledge the continued support of the Australian Research Council (ARC) through Discovery projects DP0984657, DP0664434, and DP0342844.

\label{lastpage}


\begin{thebibliography}{99}

\bibitem[Aldenius \& Johansson (2007)]{Aldenius07} Aldenius, M., \& Johansson, S.\ 2007, \aap, 467, 753

\bibitem[Alonso-Herrero et al.\ (1997)]{Alonso97} Alonso-Herrero, A., Rieke, M.~J., Rieke, G.~H., \& Ruiz, M.\ 1997, \apj, 482, 747

\bibitem[Bennert et al.(2011)]{Bennert11} Bennert, V.~N., Auger, M.~W., Treu, T., Woo, J.-H., \& Malkan, M.~A.\ 2011, \apj, 742, 107

\bibitem[Bettoni et al.\ (2003)]{Bettoni03}Bettoni, D., et al.\ 2003, \aap, 399, 869

\bibitem[Brand et al.(1988)]{Brand88} Brand, P.~W.~J.~L., Moorhouse, A., Burton, M.~G., et al.\ 1988, \apjl, 334, L103

\bibitem[Colina (1993)]{Colina93} Colina, L.\ 1993, \apj, 411, 565

\bibitem[Conselice, Gallagher \& Wyse (2001)]{Conselice01}Conselice, C.~J., Gallagher, J.~S.~III, \& Wyse, R.~F.~G. 2001, 
\aj, 122, 2281

\bibitem[Dopita et al.\ (1997)]{Dopita97}Dopita, M.~A., et al.\ 1997, \apj, 490, 202

\bibitem[Donahue et al.\ (2000)]{Donahue00}Donahue, M., et al.\ 2000, \apj, 545, 670

\bibitem[Fabian (1994)]{Fabian94}Fabian, A.~C.\ 1994, \araa, 32, 277

\bibitem[Fabian et al.\ (2000)]{Fabian00}Fabian, A.~C., et al.\ 2000, \mnras, 318, L65

\bibitem[Fabian et al.\ (2003a)]{Fabian03a}Fabian, A.~C., et al.\ 2003a, \mnras, 344, L43

\bibitem[Fabian et al.(2003)]{Fabian03b} Fabian, A.~C., Sanders, J.~S., Crawford, C.~S., et al.\ 2003b, \mnras, 344, L48 

\bibitem[Fabian et al.\ (2008)]{Fabian08} Fabian, A.~C., Johnstone, R.~M., Sanders, J.~S., et al.\ 2008, \nat, 454, 968 

\bibitem[Fabian et al.\ (2011)]{Fabian11} Fabian, A.~C., Sanders, J.~S., Williams, R.~J.~R., et al.\ 2011, \mnras, 417, 172 

\bibitem[Ferland et al.\ (2009)]{Ferland09} Ferland, G.~J., Fabian, A.~C., Hatch, N.~A., et al.\ 2009, \mnras, 392, 1475 

\bibitem[Fischer et al.(1987)]{Fischer87} Fischer, J., Smith, H.~A., Geballe, T.~R., Simon, M., \& Storey, J.~W.~V.\ 1987, \apj, 320, 667 

\bibitem[Gillmon et al.\ (2004)]{Gillmon04} Gillmon, K., Sanders, J.~S., \& Fabian, A.~C.\ 2004, \mnras, 348, 159

\bibitem[Graham et al.\ (2011)]{Graham11} Graham, A.~W., Onken, C.~A., Athanassoula, E., \& Combes, F.\ 2011, \mnras, 412, 2211 

\bibitem[Guillard et al.(2012)]{Guillard12} Guillard, P., Ogle, P.~M., Emonts, B.~H.~C., et al.\ 2012, \apj, 747, 95

\bibitem[Kawara \& Taniguchi (1993)]{Kawara93}Kawara, K., \& Taniguchi, Y.\ 1993, \apj, 410, L19

\bibitem[Krabbe et al.\ (2000)]{Krabbe00}Krabbe, A., et al.\ 2000, \aap, 354, 439

\bibitem[Levinson et al.\ (1995)]{Levinson95} Levinson, A., Laor, A., \& Vermeulen, R.~C.\ 1995, \apj, 448, 589 

\bibitem[Lim et al.\ (2008)]{Lim08} Lim, J., Ao, Y., \& Dinh-V-Trung 2008, \apj, 672, 252 

\bibitem[Lim et al.(2012)]{Lim12} Lim, J., Ohyama, Y., Chi-Hung, Y., Dinh-V-Trung, \& Shiang-Yu, W.\ 2012, \apj, 744, 112 

\bibitem[Lynds(1970)]{Lynds70}Lynds, R.\ 1970, \apj, 159, L151

\bibitem[McGregor et al.\ (2003)]{McGregor03}McGregor, P.~J., et al.\ 2003, SPIE, 4841, 1581, eds. Iye, M. \& Moorwood, A. F. M.

\bibitem[Minkowski (1957)]{Minkowski57}Minkowski, R.\ 1957, Proc. IAU Symp 4, ed. H.~C. Van de Hulst, Cambridge U.Press: Cambridge, p107

\bibitem[Mirabel, Sanders, \&  Kaz{\`{e}}s (1989)]{Mirabel89}Mirabel, I.~F., Sanders, D.~B., \& Kaz{\`{e}}s, I.\ 1989, \apj, 340, L9

\bibitem[Mouri (1994)]{Mouri94} Mouri, H.\ 1994, \apj, 427, 777 

\bibitem[Mouri, Kawara, \& Taniguchi (2000)]{Mouri00} Mouri, H., Kawara, K., \& Taniguchi, Y.\ 2000, \apj, 528, 186

\bibitem[Nenkova et al.\ (2008)]{Nenkova08} Nenkova, M., Sirocky, M.~M., Ivezi{\'c}, {\v Z}., \& Elitzur, M.\ 2008, \apj, 685, 147 

\bibitem[Nesvadba et al.(2010)]{Nesvadba10} Nesvadba, N.~P.~H., Boulanger, F., Salom{\'e}, P., et al.\ 2010, \aap, 521, A65 

\bibitem[Nesvadba et al.(2011)]{Nesvadba11} Nesvadba, N.~P.~H., Boulanger, F., Lehnert, M.~D., Guillard, P., \& Salome, P.\ 2011, \aap, 536, L5 

\bibitem[Nisini et al.\ (2002)]{Nisini02} Nisini, B., Caratti o Garatti, A., Giannini, T., \& Lorenzetti, D.\ 2002, \aap, 393, 1035

\bibitem[Ogle et al.(2007)]{Ogle07} Ogle, P., Antonucci, R., Appleton, P.~N., \& Whysong, D.\ 2007, \apj, 668, 699 

\bibitem[Ogle et al.(2010)]{Ogle10} Ogle, P., Boulanger, F., Guillard, P., et al.\ 2010, \apj, 724, 1193

\bibitem[Osterbrock, Shaw, \& Veilleux (1990)]{Osterbrock90} Osterbrock, D.~E., Shaw, R.~A., \& Veilleux, S.\ 1990, \apj, 352, 561

\bibitem[Pedlar et al.\ (1990)]{Pedlar90}Pedlar, A., et al.\ 1990, \mnras, 346, 477

\bibitem[Pesenti et al.(2003)]{Pesenti03} Pesenti, N., Dougados, C., Cabrit, S., et al.\ 2003, \aap, 410, 155 

\bibitem[Pradhan \& Zhang (1993)]{Pradhan93} Pradhan, A.~K., \& Zhang, H.~L.\ 1993, \apjl, 409, L77

\bibitem[Prieto (1996)]{Prieto96}Prieto, A.~M.\ 1996, \mnras, 282, 421

\bibitem[Quinet et al.\ (1996)]{Quinet96} Quinet, P., Le Dourneuf, M., \& Zeippen, C.~J.\ 1996, \aaps, 120, 361

\bibitem[Riffel, Rodr\'iguez-Ardila, \& Pastoriza (2006)]{Riffel06} Riffel, R., Rodr\'iguez-Ardila, A., \& Pastoriza, M.~G.\ 2006, \aap, 457, 61

\bibitem[Rodr\'iguez-Ardila, Riffel \& Pastoriza (2005)]{Rodriguez05}Rodr\'iguez-Ardila, A., Riffel, R.,  \& Pastoriza, M.~G. 2005, \mnras, 364, 1041

\bibitem[Rubin et al. (1977)]{Rubin77} Rubin, V.~C., Oort, J.~H., Ford, W.~K., Jr., \& Peterson, C.~J.\ 1977, \apj, 211, 693 

\bibitem[Rudy et al.(1993)]{Rudy93} Rudy, R.~J., Cohen, R.~D., Rossano, G.~S., et al.\ 1993, \apj, 414, 527 

\bibitem[Sabra et al.\ (2000)]{Sabra00} Sabra, B.~M., Shields, J.~C., \& Filippenko, A.~V. 2000, \apj, 545, 157

\bibitem[Salom{\'e} et al.(2006)]{Salome06} Salom{\'e}, P., Combes, F., Edge, A.~C., et al.\ 2006, \aap, 454, 437 

\bibitem[Salom{\'e} et al.(2008a)]{Salome08} Salom{\'e}, P., Revaz, Y., Combes, F., et al.\ 2008a, \aap, 483, 793 

\bibitem[Salom{\'e} et al.(2008b)]{Salome08b} Salom{\'e}, P., Combes, F., Revaz, Y., et al.\ 2008b, \aap, 484, 317 

\bibitem[Salom{\'e} et al.\ (2011)]{Salome11} Salom{\'e}, P., Combes, F., Revaz, Y., et al.\ 2011, \aap, 531, A85 

\bibitem[Sanders \& Fabian(2007)]{Sanders07} Sanders, J.~S., \& Fabian, A.~C.\ 2007, \mnras, 381, 1381 

\bibitem[Storchi-Bergmann et al.\ (2009)]{Storchi-Bergmann09} Storchi-Bergmann, T., et al.\ 2009, \mnras, 394, 1148

\bibitem[Sutherland \& Bicknell(2007)]{Sutherland07} Sutherland, R.~S., \& Bicknell, G.~V.\ 2007, \apjs, 173, 37 

\bibitem[Takami et al.(2006)]{Takami06} Takami, M., Chrysostomou, A., Ray, T.~P., et al.\ 2006, \apj, 641, 357 

\bibitem[Thompson(1995)]{Thompson95} Thompson, R.~I.\ 1995, \apj, 445, 700

\bibitem[Wagner \& Bicknell(2011)]{Wagner11} Wagner, A.~Y., \& Bicknell, G.~V.\ 2011, \apj, 728, 29 

\bibitem[Walker, Romney \& Benson (1994)]{Walker94}Walker, R.~C., Romney, J.~D., \& Benson, J.~M.\ 1994, \apj, 430, L45

\bibitem[Wilman, Edge \& Johnstone (2005)]{Wilman05}Wilman, R.~J., Edge, A.~C. \& Johnstone, R.~M.\ 2005, \mnras, 359, 755

\bibitem[Wright(2006)]{Wright06} Wright, E.~L.\ 2006, \pasp, 118, 1711 

\bibitem[Unger et al. (1990)]{Unger90} Unger, S.~W., Taylor, K., Pedlar, A., et al.\ 1990, \mnras, 242, 33P

\bibitem[Zhang \& Pradhan (1995)]{Zhang95} Zhang, H.~L., \& Pradhan, A.~Z.\ 1995, \aap, 293, 953

 \end{thebibliography}
\end{document}